\documentclass{article}
\usepackage[utf8]{inputenc}
\usepackage[a4paper,top=2.54cm,bottom=2.54cm,left=2.54cm,right=2.54cm,marginparwidth=1.75cm]{geometry}

\usepackage{amsmath}
\usepackage{amsthm}
\usepackage{amssymb} 
\usepackage{graphicx}
\usepackage{subcaption}

\usepackage[english]{babel}
\usepackage{biblatex}
\usepackage{csquotes}
\usepackage{soul}

\usepackage[usenames,dvipsnames]{xcolor}

\usepackage{hyperref}

\usepackage{multirow}

\newtheorem{theorem}{Theorem}

\usepackage{authblk}

\title{Generalized Nash Equilibrium Models for Asymmetric, Non-cooperative Games on Line Graphs: Application to Water Resource Systems}

\author[1]{Nathan T. Boyd \footnote{Corresponding author. E-mail address: nboyd1@umd.edu Postal Address: 2181 Glenn L. Martin Hall, Building 088, University of Maryland, College Park, MD 20742 (N. Boyd).}}
\author[1,2]{Steven A. Gabriel}
\author[3] {George Rest}
\author[3] {Tom Dumm}
\affil[1]{University of Maryland, College Park, Maryland, USA}
\affil[2] {Norwegian University of Science and Technology, Trondheim, Norway}
\affil[3] {Ramboll, Bowie, Maryland, USA}

\begin{document}

\maketitle

\begin{abstract}
    This paper investigates the game theory of resource-allocation situations where the “first come, first serve” heuristic creates inequitable, asymmetric benefits to the players. Specifically, this problem is formulated as a Generalized Nash Equilibrium Model where the players are arranged sequentially along a directed line graph. The goal of the model is to reduce the asymmetric benefits among the players using a policy instrument. It serves as a more realistic, alternative approach to the line-graph models considered in the cooperative game-theoretic literature. An application-oriented formulation is also developed for water resource systems. The players in this model are utilities who withdraw water and are arranged along a river basin from upstream to downstream. This model is applied to a stylized, three-node model as well as a test bed in the Duck River Basin in Tennessee, USA. Based on the results, a non-cooperative, water-release market can be an acceptable policy instrument according to metrics traditionally used in cooperative game theory. 
\\
    \bf{\ul{Keywords}}: Generalized Nash equilibrium problems, non-cooperative game theory, water resources 
\end{abstract}

\section{Introduction}
\subsection{Asymmetric Games}
In engineering-economic and other systems, asymmetric games exist if some players have distinct advantages over other players. These advantages may be structural, e.g., first-mover advantage , e.g. Stackelberg games \cite{gabriel2012complementarity} or may take the form of disproportionately higher payoffs, a greater number of strategies, or other aspects. Concerns regarding equity and welfare arise when these situations involve shared resources or infrastructure of economic, social, or environmental importance. Thus, ways to balance this asymmetry provide insight into policies to improve equity in asymmetric games.  

Games played on asymmetric networks are one important class and can naturally become a source of asymmetry between players. Specifically, the asymmetric network governs the interactions among players such that only a few neighbors are capable of influencing a given player's set of decisions \cite{parise2019variational}. If this influence is biased in one direction, then the network position of certain players may be advantageous in space, time or both. Players with these positional advantages could be "indifferent" to or even exploit the strategies of others and thereby create an asymmetric game. Stackelberg games are an important example of the latter. The difference is that in Stackelberg and leader-follower games more generally, e.g., mathematical program with equilibrium constraints (MPECs) or equilibrium problems with equilibrium constraints (EPECs)  \cite{gabriel2012complementarity}, the leaders directly take into account the actions of the followers in their decision-making to optimize their own objective functions. The followers are passive and take the leaders' decisions as given.

The study of asymmetric games in this paper concentrates on an asymmetric network that takes inspiration from river systems with multiple independent water users. Specifically, the players located on the upstream end of the river have a positional advantage manifesting as privileged access to water. Downstream users must take these decisions as given, which may result in excess flooding, inadequate water supply, or degraded water quality. Independent, conflicting water usage decisions often arise in trans-boundary river basins. Namely, these include situations where the river basin is not solely contained within one administrative boundary. This general situation is known as the river-sharing problem \cite{van2007component}.

With this example in mind, we consider a general, asymmetric game on a line-graph network where a shared resource is accessed on a "first come, first served" basis. In such a network, each player is sequentially positioned in a line on a directed network \cite{van2007component}. Excluding the two terminal-end players, each player has both an upstream and downstream neighbor. The two terminal-end players have only one upstream or downstream neighbor depending on the position. The upstream users are like leaders in a Stackelberg game, except they may not directly exploit the downstream users (e.g., are "indifferent" to followers). In this context, truly indifferent leaders are mathematically equivalent to those who are abstaining from exploiting the followers. 

Using this general model, we provide several non-cooperative game theory models for linear-graph networks of river systems. The aim is to allow the downstream players to balance this asymmetry through payments to water-release markets. Compared to other papers in this line of research, the non-cooperative approach provides more realistic modeling as compared to cooperative game theoretic ones \cite{peleg2007introduction} yet still allows for an improved system benefit as compared to the current one. There are a number of important examples of these asymmetric games in a variety of different areas. Similarities and differences between river basins and other infrastructure systems are discussed in the next section.

\subsection{Water vs. Other Infrastructure Systems \label{sec:wat_inf}}
 Water resource-related risks are closely linked to a number of on-going economic and environmental concerns and have been exacerbated in recent years. The myriad number of causes are responsible for this include rapid population growth and urbanization, increased wastewater discharges and more stringent effluent limits, a greater number of recreational users, degraded in-stream environmental habitats and landscapes, increased frequency and duration of extreme climate events, and inequitable access to clean drinking water. For instance, a study found that the drought in the Western United States is the worst in 1,200 years \cite{rott_2022}. For years, stakeholders have recognized that the future was rapidly coming into focus: tackling complex challenges requires a unified collaborative approach and cutting-edge solutions to evaluate and mitigate future risks. Furthermore, translating the flow of water into the flow of benefits is inherently challenging because of water's ubiquitous usage across municipal, agricultural, and industrial sectors. This hinders the ability to validate and address the associated inequities with regulation alone.

What makes water management and river systems in particular interesting and the focus of the application in this paper, is their relationship to commodity markets. In general, there are no widely-implemented market structures to balance the asymmetry outlined above either for water quantity or quality. For example, the doctrine of prior appropriation in the western United States grants water rights on a "first in time, first in right" basis \cite{cech2005principles}. However, this system is rigid and does little more than transfer a spatial asymmetry into a temporal asymmetry. In contrast, other infrastructure systems can have market structures/systems to balance welfare and other system-level economic or other objectives.  Consider the following examples to highlight this point.

In the electric power sector, markets in Europe and North America have several stages of decisions leading up to real time.  For example, power producers can submit day-ahead bids for production levels and prices which then help independent system operators (ISOs) to balance power supply with forecasted demand. There are also markets that balance supply and demand in near real-time or automatic adjustment in real-time as well e.g., PJM power market in the U.S. (\href{https://www.pjm.com/}{https://www.pjm.com/} or Nordpool in Europe (\href{https://www.nordpoolgroup.com/}{https://www.nordpoolgroup.com/}).  

Relative to water volumes, water resources don't operate with these levels of decision-making in part because they can store water as needed.  In power systems, in today's markets there are generally no market-scale storage assets to mitigate potential imbalances in uncertain supply (i.e., renewable) or uncertain demand.  Also, power markets allow for forward contracts as well as spot markets to be as flexibile as possible which is distinct from river-based water systems.  One aspect that is akin to balancing upstream and downstream players in power is what is called demand response.  This is temporal shifting of the consumer load (e.g., residential, industrial) to better balance supply and demand.  For example, the residents in buildings may be incentivized with payments to shift their load to hours with lower prices for overall system benefit (i.e., less need for expensive and fossil fuel-based peaking plants). In this sense, the asymmetric game is over time with the upstream players the consumers (or producers) that are paid to alter their consumption (production) schedules for temporally later consumers or producers (i.e., downstream players) \cite{conejo2010real}

In transportation, specifically traffic management, there are also mechanisms in place to better balance the asymmetry in this transport infrastructure on a real-time basis.  Consider real-time tolls that change their prices based on the volume of flow along a particular highway through the use of vehicle transponders.  In effect, drivers can decide to use the roads later if the prices are too high.  The earlier drivers in this case are the upstream players whose choice of using the tolled road can affect later, downsteam drivers.  In this case the asymmetry is over time but the earlier drivers are negatively  incentivized by much higher congestion tolls (assuming that they are driving during the busy hours) \cite{gabriel1997traffic}. 

From a water-quality perspective, the analog with power is perhaps best through carbon emissions-reduction programs like the U.S. Regional Greenhouse Gas Initiative (RGGI) \href{https://www.rggi.org/}{https://www.rggi.org/}, \cite{ruth2010strategies}. This program gives certain carbon allowances (maximum  amount of tons of carbon emissions) and it’s up to the market to balance this with policy  goals. For example, power companies that produce renewable energy or can limit their carbon emissions can generate revenue from selling their unused  allowances.  Power companies that produce too much carbon emissions  have to pay for this overage.  It seems that RGGI has done well to monetize carbon emissions reductions.  From this perspective, RGGI  relates to water quality for example, sediment or pollution reduction in river systems.  While power has such systems and markets in place, it is rarer for water systems to apply them successfully. The Virginia Nutrient Credit exchange program is a notable exception \cite{Gov_McD_2012}. Another interesting comparison between water and power is that in water systems water users along a river can act as both suppliers and consumers, which is analogous to prosumers in  energy markets.

\section{Literature Review and Contributions of This Paper}
\subsection{Literature Review}

Most if not all of the research on line-graph games has been from the framework of cooperative game theory. Brink et al. (2007) used cooperative game theory to analyze line-graph games with applications in machine sequencing games and the river-sharing problem \cite{van2007component}. Khmelnitskaya (2010) extended this work to a more general case, which considers cooperative game theory on rooted-tree and sink-tree digraphs \cite{khmelnitskaya2010values}. This structure was then used to address the river-sharing problem for more complex networks. These works demonstrate that the river-sharing problem can be generalized to a mathematically abstract setting within a game-theoretic context.

Network games from the non-cooperative game theoretic framework have been researched, but lack coverage in line-graph games. For example, Cominetti et al. (2021) formulate the "Buck Passing Game" where the players attempt to pass a chore to other players in the network to minimize individual effort \cite{cominetti2021buck}. Zhou and Chen (2018) consider sequential consumption in networks during a firm's release of a new product \cite{zhou2018optimal}. It is similar to a line-graph game but involves more sophisticated network dynamics. Parise and Ozdaglar (2019) formulate a general, variational inequality framework for network games, but do not cover line-graph games as a specific case \cite{parise2019variational}. 

Water-resource problems have been analyzed from a wide variety of both cooperative and non-cooperative game theoretic network contexts. Dinar and Hogarth (2015) presented a systematic review of game theory and water resource literature \cite{dinar2015game}. They found that much of this research was related to cooperative game theory. However, the non-cooperative models lack extensive formulations from an equilibrium programming perspective.  Bekchanov et al. (2017) reviewed over 150 papers on water economic models. They concluded that the literature poorly integrates economic equilibrium models with the underlying water resource networks \cite{bekchanov2017systematic}. Archibald and Marshall (2018) corroborate this viewpoint in their literature review. They reviewed nearly 450 papers on mathematical programming in water resources, but equilibrium programming was notably absent \cite{archibald2018review}.

Britz et al. (2013) provides a notable exception to the equilibrium programming gap in the literature. They model a stylized river basin using multiple optimization problems with equilibrium constraints \cite{britz2013modeling}. In a follow-up paper, Kuhn et al. (2014) extend the approach to a real-world case study in the Lake Naivasha Basin. Despite the uniqueness of the approach, both of these papers do not attempt to generalize it to the general non-cooperative game theoretic context. They also focus primarily on water allocation issues without considering the trade-offs between water use curtailments and water infrastructure investment. Additionally, they do not consider nuances of water balances such as the role of indirect water reuse.

\subsection{Contributions of the Current Paper}
Thus, the current work formalizes the modeling approach used in \cite{britz2013modeling} as a Generalized Nash equilibrium model for asymmetric, non-cooperative games on line graphs. It also provides a counterpoint to the cooperative game theory approach that is already well covered in the literature. The goal is to demonstrate how self-enforcing agreements are possible among players even in the context of an asymmetric game. Specifically, market structures are used to identify trading opportunities that connect high marginal benefits downstream to lower marginal costs upstream. It also considers water management decisions beyond water allocation such as the role of consumptive use, indirect water reuse, storage, and capital projects.

Furthermore, the proposed water release market is more tangible than markets based on water-allocation. In the water release market, water's scarcity and associated value is based on the physical barriers and cost of releasing additional water to the river. In contrast, water-allocation markets are based on legally increasing a water-withdrawal limitation. Thus, the value of water is derived from scarcity associated with a legal barrier. In countries such as Chile, real-world implementation of these water-allocation markets are inequitable because the judicial system has not uniformly enforced this legal barrier \cite{galaz2004stealing}.

To illustrate the approach, the model is applied to a stylized river basin as well as a case study in the Duck River basin in Tennessee, USA. We consider the role of consumptive use, indirect water reuse, storage, and capital projects in water resource systems. The purpose is to extend the application of the approach used in \cite{britz2013modeling}. Taken together, the application goal is to develop a collaborative approach to water resources management to better balance the upstream-downstream asymmetries. Our approach achieves this goal using concepts from other infrastructure markets and economic theory.

Summarizing the above discussion, the current paper makes valuable contributions versus the existing literature as follows: 1. Formalize non-cooperative games on line graphs as a counterpoint to the existing cooperative game theory literature; 2. Extend non-cooperative river basin game theory models to consider engineering-economic decisions beyond water allocation schemes, and 3. Create novel water market structures to achieve a better alignment of stakeholder interests in a river basin.

\section{General Model}
\subsection{Line-Graph Network Games \label{sec:gm_lgng}}
Games on line graphs have a structure that can be expressed mathematically. The location of player $i \in I$ on the line graph can be considered as an index of sequential positions $1,2,...,|I|$. For each player, the decisions, $x_i$, and the associated payoff function, $f^{LG}_i(x_i)$, can be quantified using an optimization model. These payoffs are generated in the context where upstream players may be indifferent and/or uninfluenced by Crucially, the feasible region, $R_i$, is a function of the state variable $S \in \{s_o,s_1,s_2,...,s_{|I|}\}$. Assuming one shared resource for simplicity, the scalar $S$ represents the state of resources shared among the various players. It is constrained to take on a value $s_i$, which represents player i's transformations to the shared resources (e.g., water releases after withdrawing from the river). 

Sequential transformations to $S$ represent the primary connections from one player's optimization model to another. The function $g^{LG}$ describes the value of the state transformations from the domain of player i's optimal decisions (i.e., $x^*_i$) and the current state of the resource (i.e., $S$). 
Starting from an initial value $s_o$, each player applies these transformations sequentially according to their network position such that player i inherits the transformation to $S$ from player i-1. This nominally represents the only form of interaction among the players. 

The following algorithm mathematically expresses the dynamics of this game:

\textit{For each} $i \in I:$
\begin{enumerate}
    \item \textit{if i = 1:} $S \leftarrow{} s_o$ ; \textit{else}: $S \leftarrow{} s_{i-1}$
    \item \textit{Solve} $\max_{x_i} f^{LG}_i(x_i) \quad s.t. \quad x_i \in R_i(S)$
    \item $s_i=g^{LG}(x^*_i,S$), $i \leftarrow i+1$
\end{enumerate}
This formulation illustrates how players at a positional disadvantage (i.e., late in the sequence) inherit the shared resource. They are completely dependent on the optimal decisions, $x^*_i$ of the players early in the sequence yet have no direct opportunity to influence them because each program is sequentially solved. Beyond understanding the magnitude of potential inequities, a solution to this system is rather trivial in the aggregate because each player effectively optimizes in isolation.
\subsection{Generalized Nash Reformulation}
The line-graph network structure does not eliminate the possibility for mutually-beneficial actions if players are allowed to interact. Such a system is reformulated in (\ref{eqn:gen_nash_opt}) where each player separately solves the following optimization problem:
\begin{subequations}
    \begin{equation}
        max_{x_i} f^{GN}_i(x_i,y_i,z_i,\pi) 
        \label{eqn:gen_nash_of} :
    \end{equation}
   \texttt{s.t.}
    \begin{equation}
          x_i,y_i,z_i \in R^{GN}(s_{i-1}(y_i))  
          \label{eqn:gen_nash_fr}
    \end{equation}
    \label{eqn:gen_nash_opt}
\end{subequations}
The vector $y_i$ represents player i's decisions that desirably influence players early in the sequence to alter their utilization of the shared resource, $S \in \{s_1, s_2, ... s_n\}$ (e.g., revenue-generating water purchases). The term $s_o$ is not included here because it is a boundary condition representing the initial unaltered state of the resource. Mathematically speaking, the vector $y_i$ indirectly shows up in the objective functions of other players not equal to $i$. Conversely, the vector $z_i$ represents player i's accommodations for players later in the sequence (e.g., additional water releases downstream). The variable $\pi$ is a vector of variables unique to the system that informs each player's decisions (e.g., market prices).  This is a generalized Nash problem as the constraint region is affected by other players' decisions.

The values of $s_i$ and $\pi$ are exogenous to each player but are endogenous to the system as shown in (\ref{eqn:gen_nash_sys1}):
    \begin{equation}
        s_i = g^{GN}(z_i,s_{i-1}), 
        \pi = h(Y,Z)
        \label{eqn:gen_nash_sys_interact}
    \end{equation}
    \label{eqn:gen_nash_sys1}
The function $g^{GN}$
describes the value for the variable $s_i$ on the domain of player i's feasible accommodation decisions, $z_i$, and the feasible states of the inherited resource, $s_{i-1}$. In the simplest form, this function could include conservation of flow constraints. In its most complex form, this function could include simulation outputs from a physical process model.  The function h transforms the vector of decisions $Y = (y_1^T, ..., y_{|I|}^T)^T$ and $Z = (z_1^T, ..., z_{|I|}^T)^T$ to some system value $\pi$ (e.g., market price).


\subsection{Cooperative Game Theoretic Considerations \label{sec:gm_coop}}
Topics from cooperative game theory are used to  compare alternative systems or rules for non-cooperative player interactions in the models below.  The first concept involves the characteric function $v(I^C)$ for each subset of players $I^C  \subseteq I$.  This function gives the amount that the members of $I^C$ are guaranteed to receive if they act together as a coalition \cite{winston2004operations}. Given the non-cooperative model structure specified above, the conditions for inclusion in $I^C$ are as follows:
\begin{equation}
    I^C = \{i|\quad ||y_i||\neq 0\} \cup \{i|\quad ||z_i|| \neq 0\}
\end{equation}
where $y_i$ and $z_i$ are the same variables defined previously and $||\cdot||$ is any vector norm.
In the line-graph game, the value for the characteristic function can be calculated as follows:
\begin{equation}
    v(I^C) = \sum_{i \in I^C}\left( f^{GN}_i(x^*_i,y^*_i,z^*_i,\pi) - f^{LG}_i(x^*_i)\right)
    \label{eqn:line_graph_cf}
\end{equation}
where $x^*_i, y^*_i,z^*_i$ represent optimal decisions for player i. Thus, the characteristic function is the sum of the improvement from the Generalized Nash reformulation over the original line-graph game across all participating players. 

The second concept is the notion of an imputation. It is defined mathematically as follows \cite{winston2004operations}:
\begin{subequations}
    \begin{equation}
        v(I) = \sum^{|I|}_{i=1} r_i
        \label{eqn:imput_cond_1},  \quad
    \end{equation}
    \begin{equation}
        r_i \ge v(\{i\}) \quad \forall i \in I
        \label{eqn:imput_cond_2}
    \end{equation}
\end{subequations}
where $r_i$ is the reward player i receives from participating in the coalition. The first condition states that the rewards distributed to all players must equal the value of the characteristic function composed of all players. The second condition states that participating in the coalition should not decrease the rewards received. Put another way, joining the coalition of the group should provide a higher reward than the coalition of only oneself. Therefore, an imputation must maximize the payoff to the coalition and leave no player worse off than they would be independently. It effectively is a condition for mutual interest among the participating players.

\begin{theorem} \label{th:imputation}
A non-negative reward $r_i$ for all players is a sufficient condition for an imputation in this optimization-based, line-graph game. 
\end{theorem}

See the Appendix for the proof.  Using these metrics, the solutions of alternative reformulations shown below are compared and contrasted using the associated characteristic functions and the criteria for imputations. Specifically, we seek alternative reformulations with high characteristic function values that are also imputations. Such conditions represent interactions that are non-cooperative fundamentally but are mutually beneficial for the players.

\section{River Basin Equilibrium Model Formulations}
This section presents a special case of the general line-graph model applied to water resources in river basins.  Two alternative Generalized Nash reformulations are considered in Subsections \ref{sec:gcm_form} and \ref{sec:csm_form}. In both cases, the interaction structure between the players (i.e., the $\pi$ function in Equation \ref{eqn:gen_nash_sys_interact}) is a water-release market. Subsection \ref{sec:no_mkt} considers the original line-graph game without any market to allow interaction between the players. Finally, Subsection \ref{sec:performance} translates the cooperative game theory concepts into performance metrics for the market structures.

\subsection{Model Overview \label{sec:rbe_overview}}

\begin{figure}
\centering
\includegraphics[scale = 0.75]{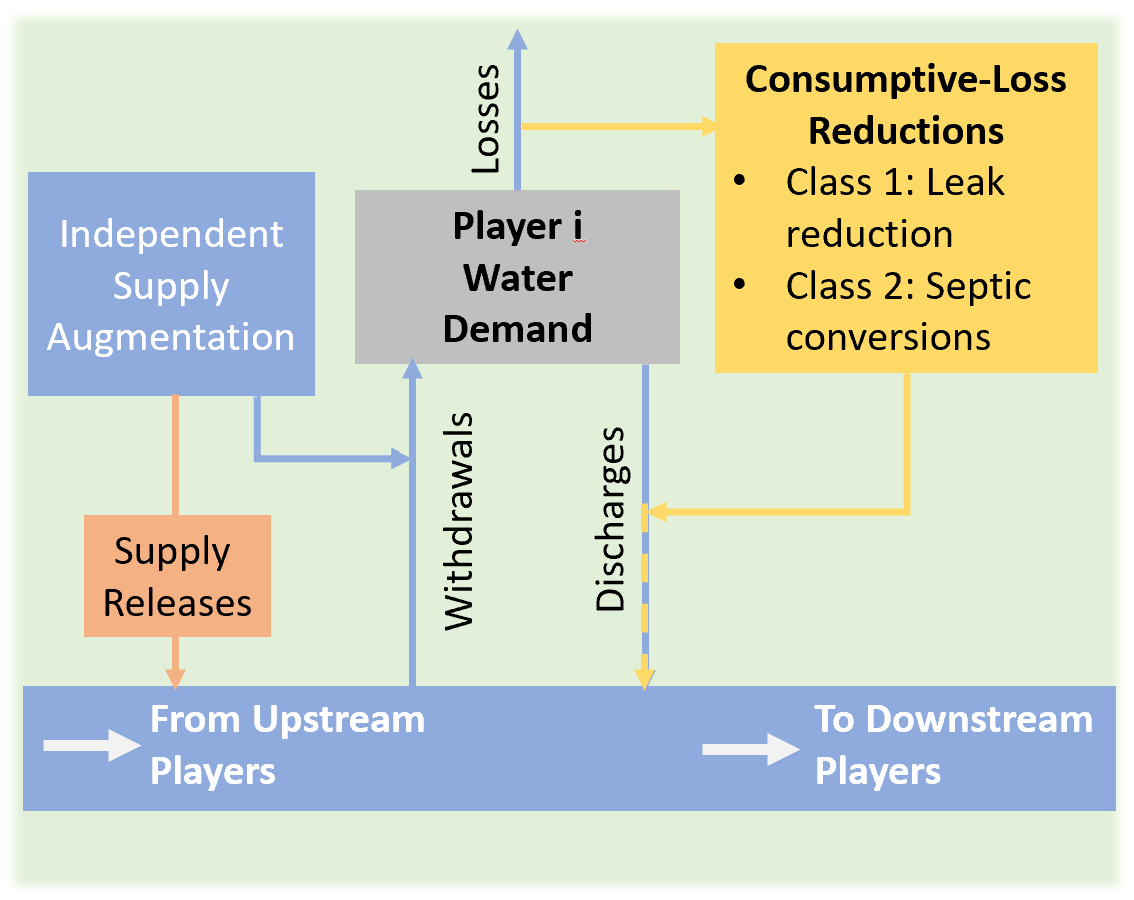}
\caption{Hydrology and flow balance in the river basin for player i.}
\label{fig:flow_schem}
\end{figure}

Figure \ref{fig:flow_schem} depicts the flow balance for a particular river user, referred to as player i. The water available to player i is a function of the flow rate in the river as well as water released from upstream players. Player i can withdraw river water from an intake and combine it with an independent water supply, which typically represents a capital project such as a reservoir or pumped groundwater. The total water withdrawn is used to meet demand. A fraction of this water usage is returned to the river as discharges in the form of runoff, wastewater, or both. It recombines with the river water and flows to downstream players. The remaining water fraction is returned to the groundwater or another basin. This fraction is called consumptive losses because they effectively are unavailable for downstream players.

Releases from independent water supply infrastructure are one mechanism to increase flows in the river. Specifically, water is released from a structure, such as a dam, into the river to increase the flow levels. It typically provides a significant capacity increase over the natural flows in the river. A player in control of such infrastructure often gains a level of independence from other players.  Thus, modeling infrastructure of this type extends beyond water allocation schemes because supply can often be increased if demand is high enough. 

Reductions of consumptive losses are another mechanism to increase flows in the river. These reductions involve alterations such that more water is returned to the river as wastewater or runoff. One of the largest classes of consumptive losses are aging infrastructure. In this case, leaks from water mains or sewers enter the groundwater. Repairing the infrastructure reduces these losses from the river. Another class of losses are septic systems in rural and suburban areas, which return treated wastewater to the groundwater. Converting septic systems to sewers increases the return flow to the river via central wastewater collection and treatment. 

The cost profiles associated with these two sources of water releases are nearly opposite. The capacity of the independent supply infrastructure is considered to be fixed, which reflects the large fixed costs of water supply infrastructure. However, the marginal costs are low and the potential releases are high. By contrast, consumptive-loss reductions are much more diffuse. For instance, many water mains would likely need to be repaired to generate a significant increase in return flows. Accordingly, the capacity of the consumptive-loss reductions are modeled as a continuous variable. To model diminishing returns, the marginal costs are considered to be progressively higher as more consumptive-loss reductions are implemented.

\begin{figure}
\centering
\includegraphics[scale = 0.75]{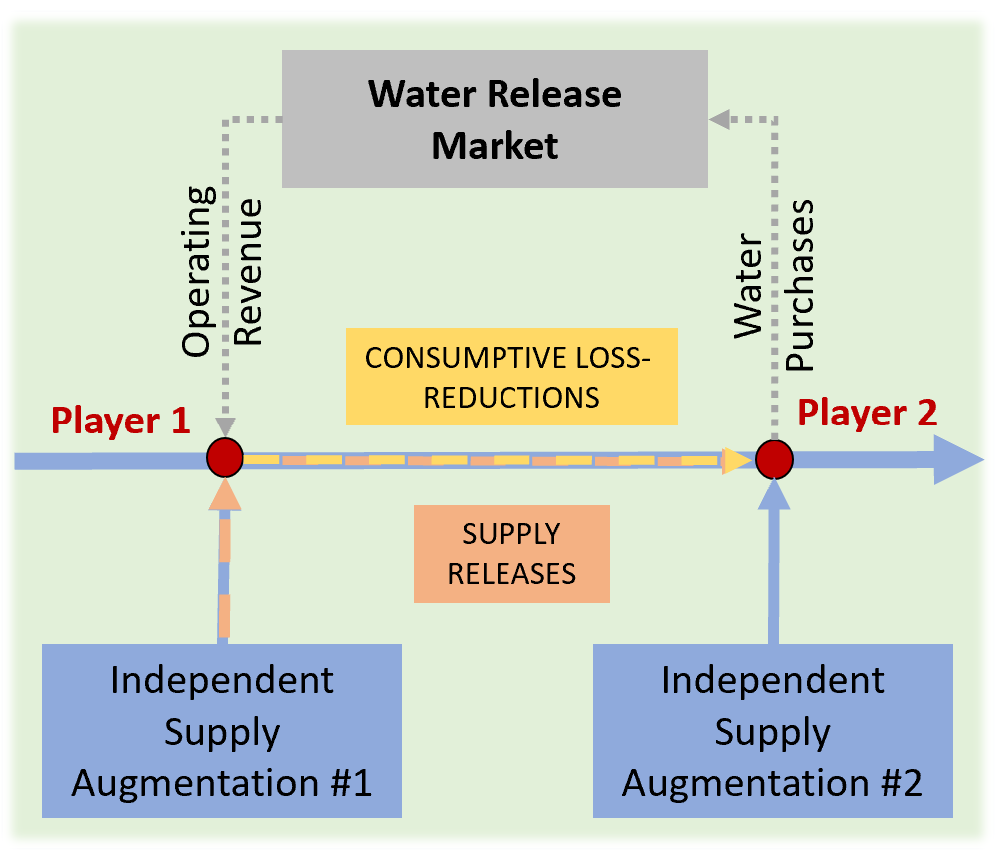}
\caption{High-level view, water-release market (just 2 players) shown. }
\label{fig:market_schem}
\end{figure}

Having considered a single player, one can envision a market structure connecting the decisions of multiple players together. Figure \ref{fig:market_schem} depicts the high-level function of the proposed water-release market. Downstream players (e.g., Player 2) purchase water from this market, which is supplied by upstream players (e.g., Player 1). This supply takes the form of water releases from independent supply sources and consumptive-loss reductions. The market prices cover the cost of these releases and could also provide additional revenue to incentivize upstream players to participate.

\subsection{Notation \label{sec:notation}}
The notation in the model consists broadly of sets, primal variables, dual variables, and parameters. In the formulation, these are either units of flow or unit costs per flow rate. In the results section, flow rates are expressed in million gallons per day (MGD), or equivalently, thousand cubic meters per day (TCMD). Notation representing unit costs are expressed in discounted million dollars/MGD/planning period \((\$M/MGD)\), or equivalently, discounted million dollars/TCMD/planning period. \((\$M/TCMD)\). These latter units were chosen to represent flow rates conventionally while allowing total costs to be calculated over longer planning periods.

\subsubsection*{Sets} 
 The following list consists of the sets in the model. Aliases are provided to allow the calculation of cumulative values arising from the use of these sets in the model. The brackets are omitted when referring to the cardinality of these sets. 
    \begin{itemize}
        \item $i, j, k \in \{I\}$ = indexed users of the river from upstream to downstream
        \item $U_i \subset I$ = upstream nodes of $i$, where $j\in U_i$ is a typical node index
        \item $D_i \subset I$ = downstream nodes of $i$, where $k\in D_i$ is a typical node index
        \item $c \in \{C\}$ = classes of water loss reductions in ascending order of expense
        \item $t, t' \in \{T\}$ = budgetary planning time periods
    \end{itemize}

\subsubsection*{Primal Variables} 

The following consists of the primal variables in each user's optimization problem. In this context, reliable capacity is a deterministic equivalent corresponding to a reasonable probability that the supply will be available to meet water demands.

    \begin{itemize}
        \item \(W^{D}_{i,t}\) = player i's incremental increase in direct water withdrawal from the river relative to time period t-1 (volume/day)
        \item \(W^{S}_{i,t}\) = player i's water supply sources from capital improvements that are independent of upstream releases (volume/day).
        \item \(Q_{i,t}\) = player i's total demand in time period t (volume/day)
        \item \(K_{i,t}\) = player i's reliable capacity added from capital project in time period t (volume/day)
        \item \(L^{R}_{i,c,t}\) = player i's incremental water loss reductions in class c in time period t (volume/day)
        \item \(W^{P}_{i,t}\) = player i's water purchases from upstream in the cost-sharing market formulation to reduce asymmetric access to water in time period t (volume/day)
        \item \(W^P_{i,j,t}\) = player i's purchases from an upstream player j (volume/day)
        \item \(W^P_{k,i,t}\) = water sales to player k downstream from i (volume/day)
        \item \(O^{min}_{i,t}\) = player i's minimum water outflow to downstream nodes in time period t (volume/day)
    \end{itemize}
\subsubsection*{Dual Variables}
    \begin{itemize}
        \item \(\gamma^{loss}_{i,c,t}\) = Nonnegative shadow price for loss reductions (cost/unit flow)
        \item \(\gamma^{flow}_{i,t}\) = Nonnegative shadow price for withdrawal limitations (cost/unit flow)
        \item \(\gamma^{cap}_{i,t}\) = Nonnegative shadow price for storage releases (cost/unit flow)
        \item \(\lambda^{sup}_{i,t}\) = Unrestricted shadow price for total supply (cost/unit flow)
        \item \(\lambda^{aug}_{i,t}\) = Unrestricted shadow price for supply augmentations (cost/unit flow)
        \item \(\lambda^{rel}_{i,t}\) = Unrestricted shadow price for the minimum release downstream (cost/unit flow)
        \item \(\pi^{as}_{i,t}\) = Nonnegative user i's price for reducing asymmetric access to water in time period t (cost/unit flow)
    \end{itemize}
\subsubsection*{Endogenous Functions} \label{subsubsect:endog_funct}
    \begin{itemize}
        \item \(B_{i,t}\) = Concave consumer benefit function for player i at time t (currency)
        \item \(R^{LR}_{i,t}\) = Revenue from loss reduction for player i at time t (currency)
        \item \(V^{op}_{i,t}\) = Total operational costs for player i at time t (currency)
        \item \(V^{inv}_{i,t}\) = Total investment costs for player i at time t (currency)
        \item \(\theta_{i,t}(Q_{i,t})\) = Endogenously defined inverse demand curve as a function of total water supply $Q_{i,t}$ (cost/unit flow)
    \end{itemize}
\subsubsection*{Parameters}
    \begin{itemize}
        \item\(c^{ops}_{i,t}\) = player i's unit operating costs in time period t (cost/unit flow)
        \item \(c^{cap}_{i,t}\) = player i's unit capital construction costs in time period t (cost/unit flow)
        \item \(c^{cu}_{i,c,t}\) = player i's unit cost for consumptive use reductions in class c during time period t (cost/unit flow)
        \item \(c^{sr}_{i,t}\) = costs incurred per release of reliable storage capacity for player i during time period t (cost/unit flow)
        \item \(d_t\) = discount rate for time period t (\%)
        \item \(\delta^{all}_{ds_{k,i}} \in \{0,1\}\) = logical parameter specifying if player k is downstream of player i
        \item \(\delta^{all}_{us_{j,i}} \in \{0,1\}\) = logical parameter specifying if player j is upstream of user i
        \item \(lf_{c,i,t}\) = player i's estimated fraction of water losses in class c and time period t (\%)
        \item \(n_{i}\) = local water inflow at player i independent of upstream players and capital improvements (volume/day)
        \item \(r^{fc}_{i,t}\) = player i's regulator imposed flow constraint in time period t (volume/day)
        \item \(a^{req}_{i,t}\) = player i's required augmentation for capital project in time period t that is varied to parameterize the equilibrium model (volume/day)
        \item \(\alpha_{i,t}\) = inverse demand intercept for player i in time period t (cost/unit flow)
        \item \(\beta_{i,t}\) = inverse demand linear slope for player i in time period t (cost/unit flow)
    \end{itemize}
    
\subsection{Formulation for Water Resource User $i\in I$ \label{sec:gen_formulation}}

Each player represents a municipal water provider who is competing with other providers for access to water along a river. Specifically, each player must decide the values for the following set of variables, $\zeta$:
\begin{align*}
\zeta=\left(W^{D}_{i,t},W^S_{i,t}, Q_{i,t},K_{i,t},L^{R}_{i,c,t},W^{P}_{i,t}  \right)  
\end{align*}
The maximum payoff and constraints for player i's decisions are modeled as an optimization problem, which is represented as (\ref{eqn:program}): 
    \begin{subequations}
        \label{eqn:program}
            \begin{align}
                \max_{\zeta} \quad 
                 \left( \sum^{|T|}_{t=1}d_t(B_{i,t} + R_{i,t}^{LR}-V^{op}_{i,t} - V^{inv}_{i,t})  \right)  
                \label{eqn:objfn}
            \end{align}
        \texttt{s.t.} \\
        \begin{equation}
            \sum^t_{t'=1}L^{R}_{c,i,t'} \le  \sum^t_{t'=1}lf_{c,i,t'}W^D_{i,t'} \quad \forall c,t
            \quad (\gamma^{loss}_{i,c,t})
            \label{eqn:const_CU}
        \end{equation}
        \begin{equation}
            Q_{i,t} \le n_{i} + W^S_{i,t} + W^P_{i,t} - r^{fc}_{i,t}+O^{min}_{i-1,t} \quad \forall t \quad (\gamma^{flow}_{i,t})
            \label{eqn:const_withdrawlim}
        \end{equation}
        \begin{equation}
            W^{S}_{i,t} \le K_{i,t} 
            \quad \forall t
            \quad (\gamma^{cap}_{i,t})
            \label{eqn:const_storrel}
        \end{equation}
        \begin{equation}
            Q_{i,t} = \sum^t_{t'=1}W^D_{i,t'} \quad \forall t \quad (\lambda^{sup}_{i,t})
            \label{eqn:supply}
        \end{equation}
        \begin{equation}
            K_{i,t} = a^{req}_{i,t}
            \quad \forall t
            \quad (\lambda^{aug}_{i,t})
            \label{eqn:const_aug}
        \end{equation}
        \begin{equation}
        W^{D}_{i,t},W^{S}_{i,t},Q_{i,t},K_{i,t},W^{P}_{i,t} \ge 0 \quad \forall t , \quad L^{R}_{i,c,t} \ge 0 \quad \forall c,t
        \label{eqn:nonneg}
        \end{equation}
    \end{subequations}


(\ref{eqn:objfn}) represents the objective function for each player. It seeks to maximize social welfare. This is measured as the discounted consumer surplus of water use and the discounted revenue from the water release market less discounted capital and operating costs. The terms in each player's objective function are defined endogenously in terms of z and are described in Section \ref{subsect:MCP Formulation}.  (\ref{eqn:const_CU}) - (\ref{eqn:nonneg}) represent the constraints on each player's objective function.  (\ref{eqn:const_CU}) states that the loss reductions cannot exceed the losses from water withdrawals.  (\ref{eqn:const_withdrawlim}) states that the water withdrawn from the river is limited to the net inflow at a particular node minus the regulatory mandated stream flow. The net inflow depends on the minimum outflow from the player immediately upstream (i.e., player i-1). Thus, individual players can modify the constraint set of another player, which results in a generalized Nash equilibrium problem \cite{facchinei2007generalized}.   (\ref{eqn:const_storrel}) states that the water released from independent water supply sources must be less than or equal to the capacity of the associated capital project.  (\ref{eqn:supply}) states that the cumulative water demand consists of the sum of incremental direct water withdrawals up to the reference time period. Equation (\ref{eqn:const_aug}) states that any capital project must be built in its entire capacity. 

\subsection{Mixed Complementarity Problem Formulation \label{sec:MCP}} 
\label{subsect:MCP Formulation}
The water equilibrium model is formulated as a mixed complementarity problem (MCP). It consists of the concatenation of every player's Karush-Kuhn-Tucker (KKT) optimality conditions associated with  (\ref{eqn:program}), the endogenous functions listed in Section \ref{subsubsect:endog_funct}, and market-clearing conditions \cite{gabriel2012complementarity}. The KKT conditions are necessary because the constraints are linear. The sufficiency direction results since each player's optimization problem (\ref{eqn:program}) is a concave maximization subject to polyhedral (hence convex) constraints.

The mixed complementarity problem generalizes the Karush-Kuhn-Tucker (KKT) optimality conditions of convex programs (with constraint qualifications), non-cooperative game theory, as well as a host of other problems in engineering and economics \cite{gabriel2012complementarity}.  Formally, the MCP is defined for a given function $F$ as finding  $x\in R^{n_x}, y \in R^{_y}$ such that
\begin{subequations} \label{eq:MCP}
\begin{align} 
0\leq F_{x}(x,y) \perp x \geq 0 \label{eq:MCPa}\\
0= F_{y}(x,y), y \texttt{ free} \label{eq:MCPb}
\end{align}
\end{subequations}
where $x$ is a non-negative vector and $y$ is a vector of free variables. The particular form of the vector-valued function $F$ is application-specific and below we describe details about this function and related optimization problems.

Two alternative water release market structures are considered within these assumptions: a general commodity market and a cost-sharing market. The asymmetric line-graph game without a market is also considered. These 
three structures result in different endogenous functions and market-clearing conditions. The similarities between them are described next, and their key features and differences are discussed in the subsections that follow.

Regardless of the market structure, there needs to be an expression defining the flow regime in the river that would result without any regulatory intervention or market. These are the flow conditions on which additional water purchases are based. 
\begin{equation}
            O^{min}_{i,t} = n_i - \sum^{|C|}_{c=1}\sum^t_{t'=1}lf_{c,i,t'}W^D_{i,t'}+\sum^{|C|}_{c=1}\sum^{t-1}_{t'=1}L^R_{c,i,t'}+O^{min}_{i-1,t} \quad \forall i,t 
            \label{eqn:min_flow}
        \end{equation}
To serve this purpose,  (\ref{eqn:min_flow}) establishes the minimum outflow conditions at each player's node. It is a function of the water withdrawal and release decisions and the minimum releases of the player immediately upstream (i.e., player i-1).

Both market structures share the same intrinsic consumer benefit function for player $i$'s water withdrawals at time $t$. As shown in  (\ref{eqn:ben_funct}), it is represented as the area under a linearized water inverse demand curve $\theta$ with intercept \(\alpha_{i,t}\) and slope $-\beta_{i,t}<0)$. 
            \begin{equation}
            B_{i,t}(Q_{i,t})=\int^{Q_{i,t}}_{0}\theta_{i,t}(x)\,dx = \int^{Q_{i,t}}_0(\alpha_{i,t}-\beta_{i,t}x\,)dx
            \label{eqn:ben_funct}
            \end{equation}
 We note that $B_{i,t}(Q_{i,t})$ is concave since, by the Leibniz Integration Rule, $\frac{d B_{i,t}(Q_{i,t})}{d Q_{i,t}}=\alpha_{i,t}-\beta_{i,t}Q_{i,t}$ so that the second derivative is just  $-\beta_{i,t}<0$.  
Both market structures also share the same function for total investment costs. Thus the total capital investment costs for node $i$ at time $t$ is expressed as follows:
            \begin{equation}
                V^{inv}_{i,t}(K_{i,t})  = c^{cap}_{i,t}K_{i,t}  
            \end{equation}

    \subsubsection{General Commodity Market (GCM) Formulation\label{sec:gcm_form}}
In the general commodity market formulation (GCM), separate markets are established for each player generating water releases. Downstream players submit bids to gain access to the water in each of these markets, and the water releases in each market are delivered to the downstream players with the highest willingness to pay at the market price. Intermediate player's between the supplier and the purchaser are not allowed to use this water when it is released into the river system. This reflects the structure of general commodity markets where the supplier establishes a price, and the supply is divided among the various consumers purchasing goods at this price.
        
A key distinction between the market structures in the way water purchases are defined. For the general commodity market, water purchases are defined as follows:
        \begin{equation}
            W^{P}_{i,t} = \sum^{|I|}_{j=1}\delta^{all}_{us_{j,i}}W_{ij,t}^{P}  \quad \forall t
            \label{eqn:trad_WP}
        \end{equation}
It states that the total water releases a recipient purchases is the sum of all the purchase requests to all upstream neighbors.
        
        The revenue from water releases for node $i$ at time $t$ is defined in terms of the price at the supplying player's market:
            \begin{equation}
                R_{i,t}^{LR}(L^{R}_{i,c,t},\forall c \in C )=\sum^{|C|}_{c=1}\pi^{as}_{i,t}(L^R_{c,i,t}+W^{S}_{i,t}) \quad \forall t
            \end{equation}
        The total operating costs are defined in terms of the marginal costs of water conveyance and treatment, supply releases, loss reductions, and water purchase requests to upstream players:
            \begin{multline}
            V^{op}_{i,t}(Q_{i,t}, L^{R}_{i,c,t},\forall c \in C,W^{S}_{i,t}, W^{P}_{ij,t},\forall j \in I, j \neq i) =    \\
            c^{ops}_{i,t}Q_{i,t} +\sum^{|C|}_{c=1}c^{cu}_{i,c,t}L^{R}_{i,c,t}+c^{sr}_{i,t}W^{S}_{i,t}+\sum_{j=1}^{|I|}\pi^{as}_{j,t}\delta^{all}_{us_{j,i}} W^{P}_{ij,t}  \quad \forall t
            \end{multline}
        
        Loss reductions are considered incremental. Namely, loss reductions only need to be made in one time period because the measures to achieve them are usually permanent (e.g., water main leak reductions). Thus, the loss reductions only need to be paid for in one time period. In contrast, releases from independent water supply sources need to be purchased in each time period, because the measures to accomplish them are not permanent.

        The market-clearing conditions say that the total amount of water demanded at node $i$ by downstream nodes to balance asymmetry  (left-hand side) should be less than or equal to what is made available by node $i$ through loss-reduction efforts or releasing from independent water supply sources.
            \begin{equation}
                \sum^{|I|}_{k=1}\delta^{all}_{ds_{k,i}} W_{ki,t}^{P} \leq  \sum_{c \in C} L^{R}_{i,c,t}+W^{S}_{i,t} \perp \pi_{i,t} \geq 0 , \forall t
                \label{eqn:gcm_mc}
            \end{equation}
        Substituting these expressions into  (\ref{eqn:program}) yields the General Commodity Market (GCM) formulation for each player:
            \begin{subequations}
            \begin{multline}
                \max_{\zeta} \quad 
                 \sum^{|T|}_{t=1}d_t(\int^{Q_{i,t}}_0(\alpha_{i,t}-\beta_{i,t}x\,)dx + \sum^{|C|}_{c=1}\pi^{as}_{i,t}(L^R_{c,i,t}+W^{S}_{i,t})-(c^{ops}_{i,t}Q_{i,t}
                 \\
                 +\sum^{|C|}_{c=1}c^{cu}_{i,c,t}L^{R}_{i,c,t}+c^{sr}_{i,t}W^{S}_{i,t}+\sum_{j=1}^{|I|}\pi^{as}_{j,t}\delta^{all}_{us_{j,i}} W^{P}_{ij,t} ) - c^{cap}_{i,t}K_{i,t})  
                 \label{eqn:gcm_of}
            \end{multline}
        \texttt{s.t.} \\
        \begin{equation} \label{eqn:gamma_loss}
            \sum^t_{t'=1}L^{R}_{c,i,t'} \le  \sum^t_{t'=1}lf_{c,i,t'}W^D_{i,t'} \quad \forall c,t
            \quad (\gamma^{loss}_{i,c,t})
        \end{equation}
        \begin{equation} \label{eqn:gamma_flow}
            Q_{i,t} \le n_{i} + W^S_{i,t} +  \sum^{|I|}_{j=1}\delta^{all}_{us_{j,i}}W_{ij,t}^{P}- r^{fc}_{i,t} + O^{min}_{i-1,t}  \quad \forall t \quad (\gamma^{flow}_{i,t})
        \end{equation}
        \begin{equation}
            W^{S}_{i,t} \le K_{i,t} 
            \quad \forall t
            \quad (\gamma^{cap}_{i,t})
        \end{equation}
        \begin{equation}
            Q_{i,t} = \sum^t_{t'=1}W^D_{i,t'} \quad \forall t \quad (\lambda^{sup}_{i,t})
        \end{equation}
        \begin{equation}
            K_{i,t} = a^{req}_{i,t}
            \quad \forall t
            \quad (\lambda^{aug}_{i,t})
        \end{equation}
        \begin{equation}
        W^{D}_{i,t},W^{S}_{i,t},Q_{i,t},K_{i,t},W^{P}_{i,t} \ge 0 \quad \forall t , \quad L^{R}_{i,c,t} \ge 0 \quad \forall c,t
        \end{equation}
    \end{subequations}
    
        The KKT conditions, endogenous functions, and market-clearing conditions are then concatenated together to form a linear complementarity problem (LCP) as shown in the Appendix. 

\subsubsection{Cost-Sharing Market (CSM) Formulation \label{sec:csm_form}}
In the cost-sharing market (CSM) formulation, the restriction on intermediate players using water releases is relaxed. This allows multiple players to claim the same quantity of released water from an upstream player. While this is uncommon in most commodity markets, direct and indirect water reuse enables water supplies to be treated as a renewable resource. Indirect water reuse is common in river basins because treated wastewater discharges feed surface water intakes downstream \cite{daniell2015understanding}. 

In contrast with the GCM structure, the markets are established at the purchasing player's node because water may be reused multiple times between the supplying and the purchasing player. Thus, the upstream players relative to the purchaser separately decide how much water releases to deliver based on the purchaser's willingness to pay. Conversely, the water release supplier receives revenue from all downstream users of this water. This market can be thought of as creating a mechanism for cost sharing through the rental of water.
    
In this structure, purchase agreements between players are no longer well defined. Therefore, water release purchases are simply the amount of water player i purchases. This makes the substitution of  (\ref{eqn:trad_WP}) into  (\ref{eqn:objfn}) and (\ref{eqn:const_withdrawlim}) unnecessary.
    
The revenue for node $i$'s water releases at time $t$ is similar to the GCM formulation, except the payment received for loss reductions is the sum of all the prices of the downstream players:
            \begin{equation}
                R_{i,t}^{LR}(L^{R}_{i,c,t},\forall c \in C, W^S_{i,t})=\sum^{|I|}_{k=1}\pi^{as}_{k,t}\delta^{all}_{ds_{k,i}}(\sum^{|C|}_{c=1}L^{R}_{i,c,t}+W^{S}_{i,t}) \quad \forall t
            \end{equation}
As before, the total operating costs are defined in terms of the marginal costs of water conveyance and treatment, supply releases, loss reductions, and water release purchases:
            \begin{equation}
            V^{op}_{i,t}(Q_{i,t}, L^{R}_{i,c,t},\forall c \in C,W^{S}_{i,t}, W^{P}_{i,t}) =    
            c^{ops}_{i,t}Q_{i,t} +\sum^{|C|}_{c=1}c^{cu}_{i,c,t}L^{R}_{i,c,t}+c^{sr}_{i,t}W^{S}_{i,t}+\pi^{as}_{i,t}W^{P}_{i,t} \quad \forall t
            \end{equation}
However, the cost for water releases is defined in terms of the price at the recipient player instead of the supplying players.
    
As in the traditional market formulation, loss reductions are permanent, and water releases from independent supply sources must be repurchased in subsequent time periods.
        
(\ref{eqn:MC}) represents the market-clearing conditions for player i's asymmetric access to water. It states that the losses reduced by the upstream players plus the amount released to the river from independent water supply sources place an upper bound on the purchases at node i. A positive price can only occur for these resources when the purchases equal the amount available upstream.

\begin{equation}
    \sum^{|I|}_{i'=1}\delta^{all}_{us_{i',i}}(\sum^{|C|}_{c=1}L^{R}_{c,i',t}+W^{S}_{i',t})-W^{P}_{i,t} \ge 0 \perp \pi^{as}_{i,t} \ge 0 \quad \forall t
    \label{eqn:MC}
\end{equation}
Substituting these expressions into  \ref{eqn:program} yields the Cost Sharing Market (CSM) formulation for each player:
            \begin{subequations}
            \begin{multline}
                \max_{\zeta} \quad 
                 \sum^{|T|}_{t=1}d_t(\int^{Q_{i,t}}_0(\alpha_{i,t}-\beta_{i,t}x\,)dx + \sum^{|I|}_{k=1}\pi^{as}_{k,t}\delta^{all}_{ds_{k,i}}(\sum^{|C|}_{c=1}L^{R}_{i,c,t}+W^{S}_{i,t})
                 \\
                 -(c^{ops}_{i,t}Q_{i,t} +\sum^{|C|}_{c=1}c^{cu}_{i,c,t}L^{R}_{i,c,t}+c^{sr}_{i,t}W^{S}_{i,t}+\pi^{as}_{i,t}W^{P}_{i,t}) - c^{cap}_{i,t}K_{i,t})  
                \label{eqn:csm_of}
            \end{multline}
        \texttt{s.t.} \\
        \begin{equation}
            \sum^t_{t'=1}L^{R}_{c,i,t'} \le  \sum^t_{t'=1}lf_{c,i,t'}W^D_{i,t'} \quad \forall c,t
            \quad (\gamma^{loss}_{i,c,t})
        \end{equation}
        \begin{equation}
            Q_{i,t} \le n_{i} + W^S_{i,t} +  W_{i,t}^{P} - r^{fc}_{i,t} + O^{min}_{i-1,t}  \quad \forall t \quad (\gamma^{flow}_{i,t})
        \end{equation}
        \begin{equation}
            W^{S}_{i,t} \le K_{i,t} 
            \quad \forall t
            \quad (\gamma^{cap}_{i,t})
        \end{equation}
        \begin{equation}
            Q_{i,t} = \sum^t_{t'=1}W^D_{i,t'} \quad \forall t \quad (\lambda^{sup}_{i,t})
        \end{equation}
        \begin{equation}
            K_{i,t} = a^{req}_{i,t}
            \quad \forall t
            \quad (\lambda^{aug}_{i,t})
        \end{equation}
        \begin{equation}
        W^{D}_{i,t},W^{S}_{i,t},Q_{i,t},K_{i,t},W^{P}_{i,t} \ge 0 \quad \forall t , \quad L^{R}_{i,c,t} \ge 0 \quad \forall c,t
        \end{equation}
    \end{subequations}
    The KKT conditions, endogenous functions, and market clearing conditions are then concatenated together to form this alternate LCP shown in the Appendix.
  
\subsubsection{No-Market Formulation \label{sec:no_mkt}}
This formulation represents no market-clearing conditions, as players only interact with each other through sequential water removal from the river. $R^{LR}_{i,t}$ is equal to zero because there is no water release market to generate revenue. Therefore, there is no incentive to reduce losses and no ability to make water purchases. The total operating costs are simplified accordingly:
    \begin{equation}
        V^{op}_{i,t}(Q_{i,t},W^{S}_{i,t}) =   c^{ops}_{i,t}Q_{i,t}+c^{sr}_{i,t}W^{S}_{i,t}\quad \forall t
    \end{equation}
Substituting these expressions into (\ref{eqn:program}) yields the No-Market formulation for each player i:
            \begin{subequations}
            \begin{equation}
                \max_{\zeta} \quad 
                 \sum^{|T|}_{t=1}d_t(\int^{Q_{i,t}}_0(\alpha_{i,t}-\beta_{i,t}x\,)dx
                 -(c^{ops}_{i,t}Q_{i,t} +c^{sr}_{i,t}W^{S}_{i,t}) - c^{cap}_{i,t}K_{i,t})  
                \label{eqn:nm_of}
            \end{equation}
        \texttt{s.t.} \\
        \begin{equation}
            Q_{i,t} \le n_{i} + W^S_{i,t} - r^{fc}_{i,t} + O^{min}_{i-1,t}  \quad \forall t \quad (\gamma^{flow}_{i,t})
        \end{equation}
        \begin{equation}
            W^{S}_{i,t} \le K_{i,t} 
            \quad \forall t
            \quad (\gamma^{cap}_{i,t})
        \end{equation}
        \begin{equation}
            Q_{i,t} = \sum^t_{t'=1}W^D_{i,t'} \quad \forall t \quad (\lambda^{sup}_{i,t})
        \end{equation}
        \begin{equation}
            K_{i,t} = a^{req}_{i,t}
            \quad \forall t
            \quad (\lambda^{aug}_{i,t})
        \end{equation}
        \begin{equation}
        W^{D}_{i,t},W^{S}_{i,t},Q_{i,t},K_{i,t}, \ge 0 \quad \forall t 
        \end{equation}
    \end{subequations}
Solving the No-Market formulation with an LCP is not necessary. It can be solved recursively using the algorithm presented in Section \ref{sec:gm_lgng}. However, the LCP Formulation is presented to be consistent with the other two Generalized Nash reformulations and is shown in the Appendix.
    
\subsection{Performance Metrics \label{sec:performance}}
The cooperative game theoretic concepts described in Section \ref{sec:gm_coop} are used to create performance metrics for the GCM and CSM market structures. They are calculated from the optimal objective function values determined from the LCP solutions, which represent social welfare. The ultimate goal of the analysis is to test the effectiveness of each market structure in reducing the asymmetry between players.

The optimal objective function values are related to the rewards, $r_i$, that each player achieves. Let $f^m_i$ represent the optimal objective function value for player i under market structure $m \in \{GSM, CSM\}$. These correspond to  (\ref{eqn:gcm_of}) and (\ref{eqn:csm_of}), which are both special cases of the generalized Nash reformulation objective function in  (\ref{eqn:gen_nash_of}). Additionally, let $f_{o_i}$ represent the optimal objective function value for player i in the no-market formulation. It is a special case of $f^{LG}_i(x_i)$ described in Section \ref{sec:gm_lgng}.

With these definitions in mind, the performance metrics from cooperative game theory can be expressed mathematically. The reward $r^m_i$ each player experiences from participating in market structure m is the difference between the player's objective function value in market structure m and the no-market case:
\begin{equation}
    r^m_i = f^m_i - f_{o_i}
\end{equation}
Accordingly, the characteristic function for market structure m, $v^m(I)$, is simply the sum of the rewards across all players:
\begin{equation}
    v^m(I) = \sum_{i \in I} r^m_i
\end{equation}

The best performing market structures will be imputations and have large characteristic function definitions. With this in mind, consider a final metric representing the difference in characteristic function values between market structures $m_1$ and $m_2$:
\begin{equation}
    v^{\Delta}(I) = |v^{m_1}-v^{m_2}|
\end{equation}
These metrics are used to analyze the numerical results presented in Section \ref{sec:results}.

\section{Results} \label{sec:results}
In this section, we discuss some theoretical results for the models proposed above with a focus on the GCM formulation for specificity. Additionally, we provide sensitivity results for a small stylized water system and numerical results for the Duck River in Tennessee, in the southeastern U.S. using real and realistic data to demonstrate insights of the models.
\subsection{Theoretical Results} \label{sec:theo_res}

These theoretical results address existence and uniqueness of solutions as well as the relationships among important variables. As will be explained, the existence of solutions can be guaranteed in certain cases. In terms of uniqueness, there is the possibility for multiple solutions to the River Basin Equilibrium Model Formulation. An example of one of the multiple solutions is provided in Section \ref{sec:3node_num}.

There are many variations in the water supply in solutions to the GCM model. For example, these variations could include more water from loss-reduction markets, water from extra supply based on capacity expansions, or decrease of demand. Accordingly, we single out a simpler yet illustrative case of what could result. Thus, the value of this illustrative case is the data input values that lead to one of these solutions in one case. It also illustrates the use of loss-reduction markets.

In this representative case, we take just one time period $|T|=1$, 1 loss-reduction class for each node, $|C|=1$ and adjust the optimization model that each node is trying to solve appropriately.  Note that there is now no discount factor (i.e., $d_t=1$), $L^R_i$ has no index for class nor time, and the incremental additions $W^D_{it}$ are now equal to $Q_i$ so that equation from before relating them is now suppressed.  Also, since the model is for the short-term, there are no capacity decisions $K_{it}, W^S_{i,t}$ are no longer needed as well as the associated dual variables to the corresponding constraints.  
\begin{subequations} \label{eq:short}
            \begin{equation}
                \max_{\zeta} \quad 
                 \int^{Q_{i}}_0(\alpha_{i}-\beta_{i}x\,)dx + \pi^{as}_{i}L^R_{i}-(c^{ops}_{i}Q_{i}+c^{cu}_{i}L^{R}_{i}+\sum_{j \in U_i}\pi^{as}_{j} W^{P}_{ij} )   
                 \label{eqn:gcm_of_short}
            \end{equation}
        \texttt{s.t.} \\
        \begin{equation} \label{eqn:gamma_loss_short}
            L^{R}_{i} \le  lf_{i}Q_i 
            \quad (\gamma^{loss}_{i})
        \end{equation}
        \begin{equation} \label{eqn:gamma_flow_short}
            Q_{i} \le n_{i} +   \sum_{j \in U_i}W_{ij}^{P}- r^{fc}_{i} + O^{min}_{i-1}  \quad (\gamma^{flow}_{i})
        \end{equation}
        \begin{equation}
        Q_{i},L^R_{i},W^{P}_{i,j}, \forall j \in U_i \ge 0  
        \end{equation}
    \end{subequations}

In addition, there are the definitional and market-clearing constraints from before suitably modified (e.g., the loss reductions happen and are used in the same time period):
\begin{subequations} \label{eq:extra_short}
\begin{equation} 
           O^{min}_{i} = n_{i}-lf_{i}Q_{i}+O^{min}_{i-1} 
           \label{eqn:omin_short}
        \end{equation}
        
        \begin{equation} 
             \sum_{k \in D_i} W_{ki}^{P} \leq  L^{R}_{i} \perp \pi^{as}_{i} \geq 0 \label{eq:MCC_short}
        \end{equation}
\end{subequations}

The next result is about existence of a solution for this one-period model (\ref{eq:short}) $\forall i \in I $ plus (\ref{eq:extra_short}) and is presented below.  For simplicity of illustration, we assume that there is just one upstream node $u$ that delivers loss reductions, one downstream node $d$ that receives it, and all other nodes $e$ are inactive relative to the loss-reduction market.  Clearly, other conditions on the data may lead to other solutions as this equilibrium problem has multiple solutions in general.

\begin{theorem} \label{eq:existence}
Consider the river system linear complementarity problem problem defined by the KKT conditions to (\ref{eq:short}) combined with (\ref{eq:extra_short}) $\forall i \in I $.  This problem always has a solution as long as the following conditions hold (assuming all positive cost coefficients):
\begin{description}\label{xxx}
\item [(i)] $\beta_i >0, \forall i \in I$
\item [(ii)] $c^{ops}_e -\alpha_e \geq 0$ for all nodes $e$ inactive in the loss-reduction market
\item [(iii)] $lf_u Q_u=W^p_{du}$
\item [(iv)] $\sum_{r=1}^e n_r \geq r^{fc}_e$ for all inactive nodes $e$. 
\item [(v)] $0<\frac{\alpha_u-c^{ops}_u}{\beta_u} \leq \sum_{r=1}^u n_r -r^{fc}_u$ for loss-reduction supplying node $u$.
\item [(vi)]$c^{cu}_u \geq \alpha_d-c^{ops}_d$ 
\item [(vii)]$\sum_{r=1}^d n_r-r^{fc}_d =-\frac{\alpha_u-c^{ops}_u}{\beta_u}lf_u<0$ 
\end{description}
where $u, d, e$ are respectively, the one loss-reduction market supplying node, the one loss-reduction receiving node, and any other node $e$ not active in the loss-reduction market.
\end{theorem}
See the Appendix for the proof.\\
Note that in (i) $\beta_i>0$ means that the demand function for each node $i$ has a strictly decreasing slope which is quite reasonable.  Also, the condition in (i) on the costs being nonnegative is also quite realistic. Condition (ii) amounts to saying that the operational costs for inactive node $e$ exceed the highest demand $\alpha_e$.  Condition (iii) says that node $d \in D_u$ uses up 100\% of the loss reductions from node $u$ that is supplying it. Condition (iv) states that the inflows from upstream nodes is sufficient to cover the required flow constraints for each nodes $e$ that are inactive in the loss-reduction market. Condition (v) states that the extra supply at node $u$ after accounting for inflows and flow constraints should be sufficiently large.  Condition (vi)  is a statement connecting the loss-reduction costs for upstream node $u$ with the largest demand and operational costs for downstream node $d$.  Lastly, (vii) states that the inflow to node $d$ from all upstream nodes and itself less any flow restrictions should be sufficiently negative to induce water purchases from upstream node $u$. 

We now return to the original GCM model allowing for multiple time periods and loss-reduction classes. With this in mind, the next result concerns the prices at active nodes for the GCM model.  Here we adopt the following notation:  $U_{it}^+$ is the set of nodes upstream of node $i$ where node $i$ purchases loss-reduction measures at time $t$ from node $j$, i.e., $U_{it}^+=\{j \in I: j\in U_i, W^P_{ijt}>0\}$.  Also,   $D_{it}^+$ is the set of nodes downstream of node $i$ where node $i$ sells loss-reduction measures at time $t$ to node $k$, i.e., $D_{it}^+=\{k\in D_i:  W^P_{kit}>0\}$.  The result below implies that the upstream loss-reduction markets must equilibrate in prices so that there is no arbitrage opportunity among them relative to a fixed, downstream node $i$ that wants these loss-reduction measures. 
\begin{theorem} \label{th:prices}
Consider node $i$ in the GCM formulation and all upstream nodes $j \in U_{it}^+$.   Then, $\pi^{as}_{jt}=\pi_{it}, \forall j \in U_{it}^+$ where $\pi_{it}$ is a common loss-reduction market price for these nodes $j \in U_i$.
\end{theorem}
See the Appendix for the proof.

The next result is a statement about the uniqueness of the nodal prices for a fixed value of the dual variable $\gamma^{loss}_{i,c,t}$.  This dual variable is associated with constraint (\ref{eqn:gamma_loss}) and it can be construed as the incremental value of one more unit of loss reductions.  The first part of the next result (\ref{eqn:part a}) states that the prices at a node $i$ are the (discounted) sum of future values of  $\gamma^{loss}_{i,c,t}$ plus the unit cost for consumptive use class $c$. This can be understood as the sum of the future opportunity costs plus the current operating cost of loss reductions.  The second part of this next result (\ref{eqn:part b})says that these nodal prices also are the discounted shadow price for downstream nodes $k \in D_{it}^+$ of getting one more unit of flow via the associated multiplier $\gamma^{flow}_{kt}$ to constraint (\ref{eqn:gamma_flow}).  Thus, these nodal prices are computed to balance the economic needs of node $i$ with its downstream loss-reduction market customers.

\begin{theorem} \label{th:prices2}
Consider the GCM formulation and a node $i$ for which there are loss reductions, i.e., $L^R_{c,i,t} >0$ for some $c \in C, t \in T$.  Then, 
\begin{subequations} 
\begin{equation} \label{eqn:part a}
 \pi^{as}_{it} =\frac{\sum_{t'=t} ^{|T|} \gamma^{loss}_{i,c,t}}{d_t}+c^{cu}_{i,c,t}   
\end{equation}
\begin{equation} \label{eqn:part b}
 \pi^{as}_{it}=\frac{\gamma^{flow}_{k,t}}{d_t}, \forall k \in D_{it}^+    
\end{equation}
\end{subequations}
\end{theorem}
See proof in the Appendix.

Given that there are multiple solutions to the river basin equilibrium problem as indicated above, some other approach is needed perhaps to further refine the solution set.  For example, one approach is to use equity-enforcing constraints or more generally logical constraints to filter out all but one solution as presented in the discretely constrained mixed complementarity problem (DC-MCP) formulations from \cite{gabriel2017solving, djeumou2019applications}.

\subsection{Numerical Results}
\subsubsection{Three-Node Model}
\label{sec:3node_num}
A  small but illustrative three-node model of the above two formulations was developed to illustrate the merits of the modeling approach and to compare the market structures. Three players, numbered sequentially in increasing order from upstream to downstream, represent producers, prosumers, and consumers of consumptive loss reductions, respectively. Two loss-reduction classes and two time periods give the model multi-dimensionality with respect to these parameters. In contrast, no capital projects are present to simplify the analysis.

The parameters for this system were chosen to represent a system with asymmetric access to water and intermediate water scarcity. Enough water is available in the first time period for two out of three players to completely satisfy their demand. In the second time period, the economic growth potential of all players is greater than the water available. To ensure asymmetry, all of the inflow into the river basin occurs at Player 1's node ($n_1$). Each player has the same regulatory flow constraint ($r^{f,c}_{i,t}$) representing the minimum base flow necessary to preserve the aquatic habitat. The overall intent is to fully utilize the market in time period 2 while illustrating reasonable starting conditions in time period 1.

Within these guidelines, various scenarios were developed to consider different types of river basins. Certain parameters were kept constant from scenario to scenario to keep them relatively comparable. These include each player's operating costs ($c^{ops}_{i,t}$), the discount rate ($d_t$), and the inverse demand slope ($\beta_{i,t}$). The remaining parameters were varied across scenarios to ascertain the impact of different player configurations, supplies, and demands. These include the following four non-fixed parameters: maximum demand in time period 1, the maximum demand in time period 2, the loss fractions ($lf_{c,i,t}$), and the consumptive use reduction costs ($c^{cu}_{i,c,t}$). 

The demands were incorporated into the model via the inverse demand intercept ($\alpha_{i,t}$). Specifically, a point of known price and quantity was assumed to exist at the current demand level and operating cost ($c^{ops}_{i,t}$). Linearizing about this point for a given value of $\beta_{i,t}$ was then used to obtain the intercept value.  \ref{eqn:alpha_linear} expresses this mathematically:
\begin{equation}
    \alpha_{i,t} = \beta_{i,t}demand_{i,t}+c^{ops}_{i,t}
    \label{eqn:alpha_linear}
\end{equation}

Table \ref{tab:toy_model_data} summarizes the input data used in the three-node model scenarios. Each row represents the unique value assigned to the parameter from the first column. The second, third, and fourth columns indicate the applicable indices for the given value assignment. For example, the inverse demand slope ($\beta$) is assigned a value of 3.0 for all three players in all three time periods. Consumptive loss-reduction costs, time period 1 demands, time period 2 demands, and loss factors are baseline values to be varied in the different scenarios \footnote{For simplicity, only the baseline values themselves are depicted.}. 

\begin{table}
\centering
\begin{tabular}{ccccc}
Parameter                & Player(s) & Time Period(s) & Class(es) & Value \\ \hline
int rate                      & 1, 2, 3   & 1, 2           &           & 0.04  \\ \hline
$\beta$                  & 1, 2, 3   & 1, 2           &           & 3     \\ \hline
\multirow{2}{*}{n}       & 1         &                &           & 9     \\
                         & 2, 3      &                &           & 0     \\ \hline
rfc                      & 1, 2, 3   &                &           & 4     \\ \hline
c\_ops                   & 1, 2, 3   &                &           & 1     \\ \hline
\multirow{2}{*}{c\_cu\footnotemark[\value{footnote}]}  & 1, 2, 3   &                & 1         & 1     \\
                         & 1, 2, 3   &                & 2         & 5     \\ \hline
\multirow{2}{*}{demand\footnotemark[\value{footnote}]} & 1, 2, 3   & 1              &           & 5     \\
                         & 1, 2, 3   & 2              &           & 10    \\ \hline
lf\footnotemark[\value{footnote}]                      & 1, 2, 3   & 1, 2           & 1, 2      & 0.1   \\ \hline
\end{tabular}
\caption{Data used in the three-node model scenario analysis.}
\label{tab:toy_model_data}
\end{table}

The baseline values, which are the last three parameters in Table \ref{tab:toy_model_data}, are systematically varied among the players to investigate how player heterogeneity influences market structure. For a given scenario, the baseline values are each multiplied by a low (0.66), medium (1.0), or high (1.33) scaling factor to obtain the actual value of the parameter. To create this heterogeneity, no two players have the same scaling factor for a given parameter in a given scenario. Mathematically, this results in 3! = 6 ways to assign a player a value for a given parameter. For example, consider one of the six ways to assign the time period 2 demands among the players. The scaling factors for players (1, 2, 3) are (medium, high, low) respectively such that the tuple of associated demand values is (10.00,13.33,6.66).

A given scenario consists of one of these player-specific tuples for each of the four baseline value parameters. Every combination of tuples for the four parameters were considered, which results in a sizable number of scenarios. Mathematically, applying the Cartesian product to the four non-fixed parameters results in  $3!^4 = 1296$ scenarios total. Table \ref{tab:data_detailed} provides examples of final parameter values that correspond to scenarios analyzed closely in this section.

Each scenario was compiled and solved in GAMS \footnote{ www.gams.com/} using an application programming interface with Python. Three separate models were solved for each scenario, which included the GCM, CSM, and no-market formulations. The solutions from these models were used to calculate the metrics detailed in Section \ref{sec:performance}. These are used to quantify the benefit of the GCM and CSM market structures relative to no market.

Table \ref{tab:toy_summary_data} summarizes the results of the various scenario runs in more detail. The first row depicts the percentage of the scenarios where the market structure in the corresponding column generated higher social welfare improvements (i.e. $v^m(N)$) than the alternative (GCM or CSM). The second and third row depict the mean and standard deviation, respectively, of social welfare improvement as defined in Equation \ref{eqn:objfn}. Lastly, the fourth and fifth rows depict the improvement of social welfare when the given market structure is higher-performing (i.e., $v^{\Delta}(N)$).

\begin{table}
\centering
\begin{tabular}{l|ll}
Metric                         & GCM     & CSM     \\ \hline
\%  higher $v^m(N)$              & 3.70\%  & 96.30\% \\
average $v^m(N)$                  & \$42.64 & \$36.87 \\
std. dev $v^m(N)$                 & \$0.74  & \$11.96 \\
average $v^{\Delta}(N)$               & \$42.64 & \$14.49 \\
std. dev $v^{\Delta}(N)$              & \$0.74  & \$14.44
\end{tabular}
\caption{Summary of Scenario Results}
\label{tab:toy_summary_data}
\end{table}

For every player in every scenario, at least one of the market structures improves social welfare as compared to no market at all (i.e., $r_i \ge 0$).  Therefore, the markets demonstrate mutual interest among the players (i.e., always form an imputation). However, not all the market structures result in mutually beneficial interactions. For instance, in some of the scenarios, the lower-performing market structure will not form an imputation. This illustrates why choosing the proper market structure is important. 

The CSM generates higher social welfare for most of the scenarios compared to the GCM. This occurs in over 96 \% of the scenarios analyzed. However, the GCM vastly outperforms in this metric relative to the CSM in the small number of its preferred scenarios. It reliably generates high social welfare improvements when it is the appropriate choice. The average and standard deviations for $v^m(N)$ and $v^{\Delta}(N)$ in Table \ref{tab:toy_summary_data} support this finding.

The furthest downstream player always undergoes dramatic economic growth for scenarios where the GCM has higher social welfare than the CSM. Specifically, Player 3 has a low demand in time period 1 and grows to have a high demand in time period 2. Under these conditions, Player 3 is most vulnerable to the decisions of Player 2. As will be explained, this vulnerability exists when the system-wide social welfare is most sensitive to the positional advantage of intermediate players along the network. This demand profile appears in all the cases where the GCM exceeds the CSM in social welfare. 

The loss-reduction market, regardless of structure, is most effective relative to no market for a specific set of common scenarios. These occur when the time period 2 demands get progressively higher downstream and the losses get progressively higher upstream. This finding makes sense, because higher losses upstream lead to less water available downstream where it is most valuable. Interestingly, these scenarios were relatively insensitive to the consumptive use costs. The demand and loss parameters seem to matter the most to the viability of the market.

With the aggregate results in mind, three scenarios were selected for detailed analysis. The objective was to identify a scenario for each market structure where the social welfare differences are the most pronounced. Mathematically, these are scenarios where $v^m(N)$ and $v^{\Delta}(N)$ are both high. The first scenario, hereafter called "Large Prosumer Scenario," was chosen to illustrate the merits of the CSM market. The second scenario, hereafter called "Downstream Economic Growth Scenario," was chosen to illustrate the merits of the GCM market. The parameter values used in these scenarios are shown in Table \ref{tab:data_detailed}. This further reinforces the importance of the demand profile to the appropriate water-release market structure. Furthermore, a third scenario was selected to illustrate the non-uniqueness of prices as described in Section \ref{sec:theo_res}.

\begin{table}
\centering
\begin{tabular}{|l|l|lll|}
\hline
\multirow{2}{*}{Parameter}      & \multirow{2}{*}{Scenario} & \multicolumn{3}{l|}{Player}                                     \\ \cline{3-5} 
                                &                           & \multicolumn{1}{l|}{i = 1} & \multicolumn{1}{l|}{i = 2} & i = 3 \\ \hline
\multirow{3}{*}{$demand_{i,1}$} & Large Prosumer            & \multicolumn{1}{l|}{3.33}  & \multicolumn{1}{l|}{6.67}  & 5.00  \\ \cline{2-5} 
                                & D.S. Econ. Growth         & \multicolumn{1}{l|}{5.00}  & \multicolumn{1}{l|}{6.67}  & 3.33  \\ \cline{2-5} 
                                & Multiple Prices           & \multicolumn{1}{l|}{5.00}  & \multicolumn{1}{l|}{3.33}  & 6.67  \\ \hline
\multirow{3}{*}{$demand_{i,2}$} & Large Prosumer            & \multicolumn{1}{l|}{6.67}  & \multicolumn{1}{l|}{13.33} & 10.00 \\ \cline{2-5} 
                                & D.S. Econ. Growth         & \multicolumn{1}{l|}{6.67}  & \multicolumn{1}{l|}{10.00} & 13.33 \\ \cline{2-5} 
                                & Multiple Prices           & \multicolumn{1}{l|}{10.00}    & \multicolumn{1}{l|}{13.33} & 6.67  \\ \hline
$c^{cu}_{i,1}$                  & All Three                 & \multicolumn{1}{l|}{0.67}  & \multicolumn{1}{l|}{1.00}  & 1.33  \\ \hline
$c^{cu}_{i,2}$                  & All Three                 & \multicolumn{1}{l|}{3.33}  & \multicolumn{1}{l|}{5.00}  & 6.67  \\ \hline
$lf_{1,i,1}$                    & All Three                 & \multicolumn{1}{l|}{0.13}  & \multicolumn{1}{l|}{0.10}  & 0.07  \\ \hline
$lf_{1,i,2}$                    & All Three                 & \multicolumn{1}{l|}{0.13}  & \multicolumn{1}{l|}{0.10}  & 0.07  \\ \hline
$lf_{2,i,1}$                    & All Three                 & \multicolumn{1}{l|}{0.13}  & \multicolumn{1}{l|}{0.10}  & 0.07  \\ \hline
$lf_{2,i,2}$                    & All Three                 & \multicolumn{1}{l|}{0.13}  & \multicolumn{1}{l|}{0.10}  & 0.07  \\ \hline
\end{tabular}
\caption{Parameter values used in the detailed analysis}
\label{tab:data_detailed}
\end{table}

\begin{figure}[htbp]
    \centering
    \begin{subfigure}[b]{0.4\paperwidth}
        \includegraphics[width = \linewidth]{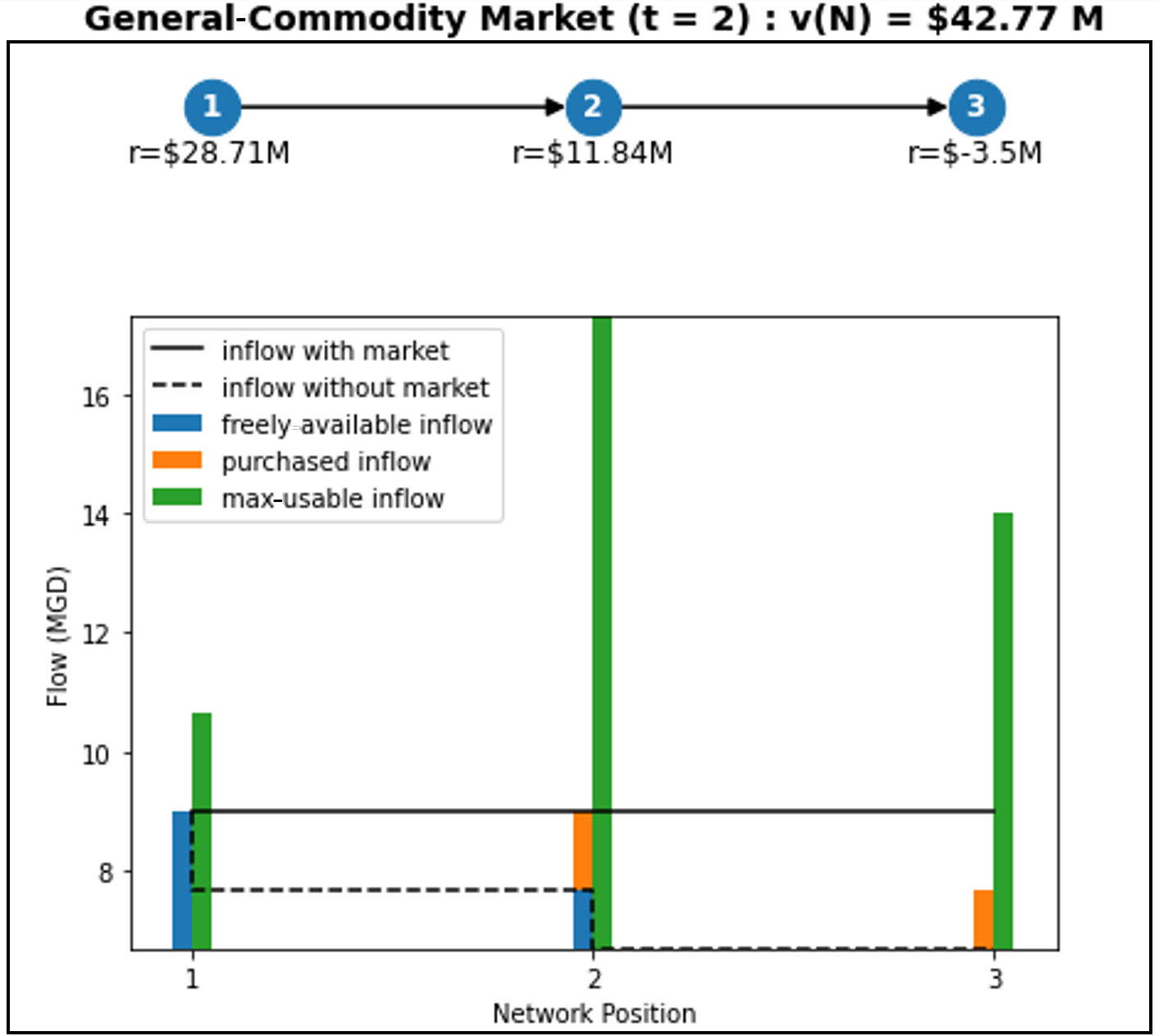}
        \caption{General commodity market with resource under-utilization.}
        \label{fig:S47_GCM}
    \end{subfigure}
    \begin{subfigure}[b]{0.4\paperwidth}
        \includegraphics[width=\linewidth]{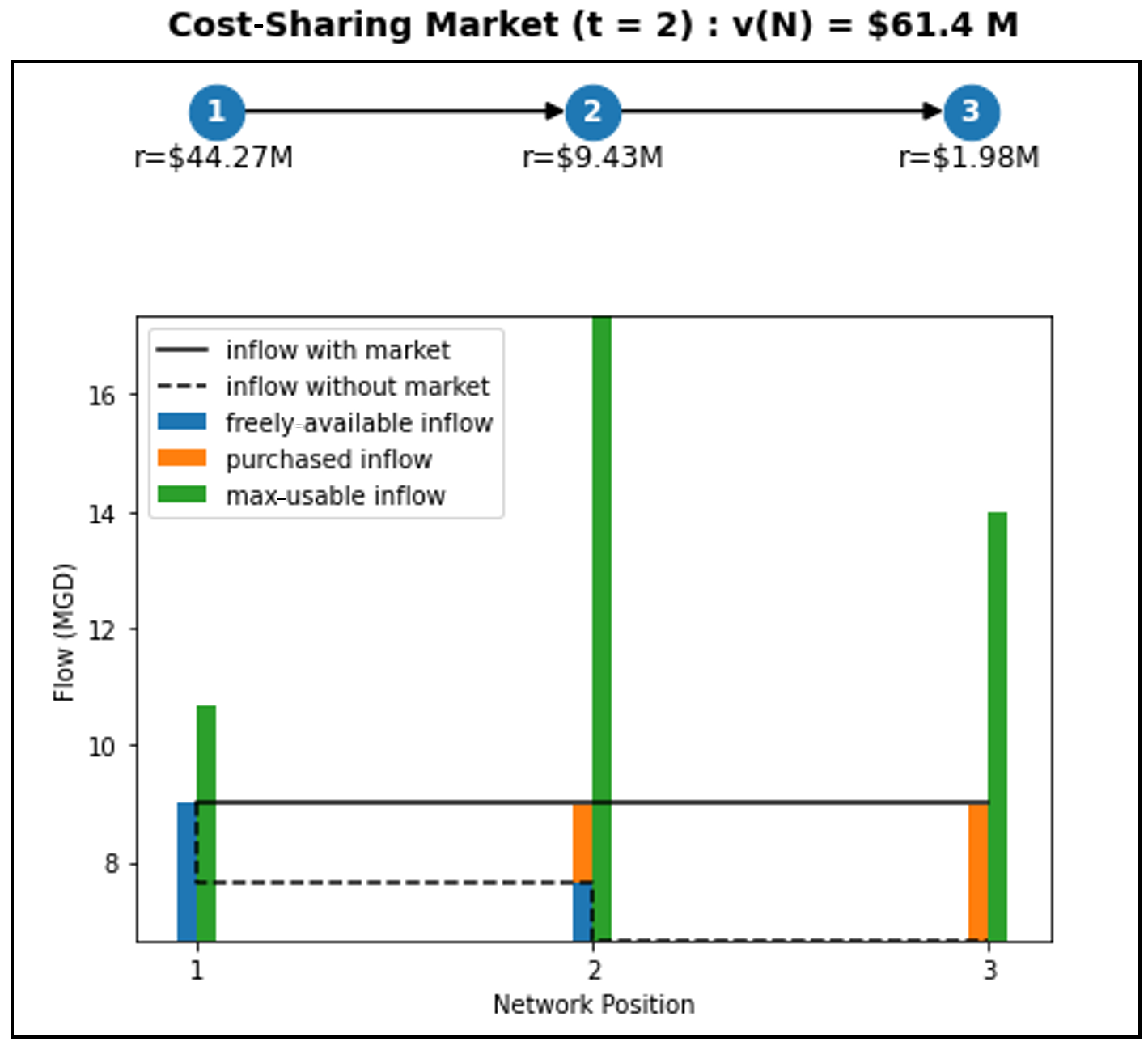}
        \caption{Cost sharing market.}
        \label{fig:S47_CSM}
    \end{subfigure}
    \label{fig:S47main}
    \caption{Visualization of the Large Prosumer Scenario. The cooperative game-theoretic rewards to each player, r, are shown in the plan view of the network. Water usage is shown in the profile view.}
\end{figure}

\begin{figure}[htbp]
    \centering
    \begin{subfigure}[b]{0.4\paperwidth}
        \centering
        \includegraphics[width=\linewidth]{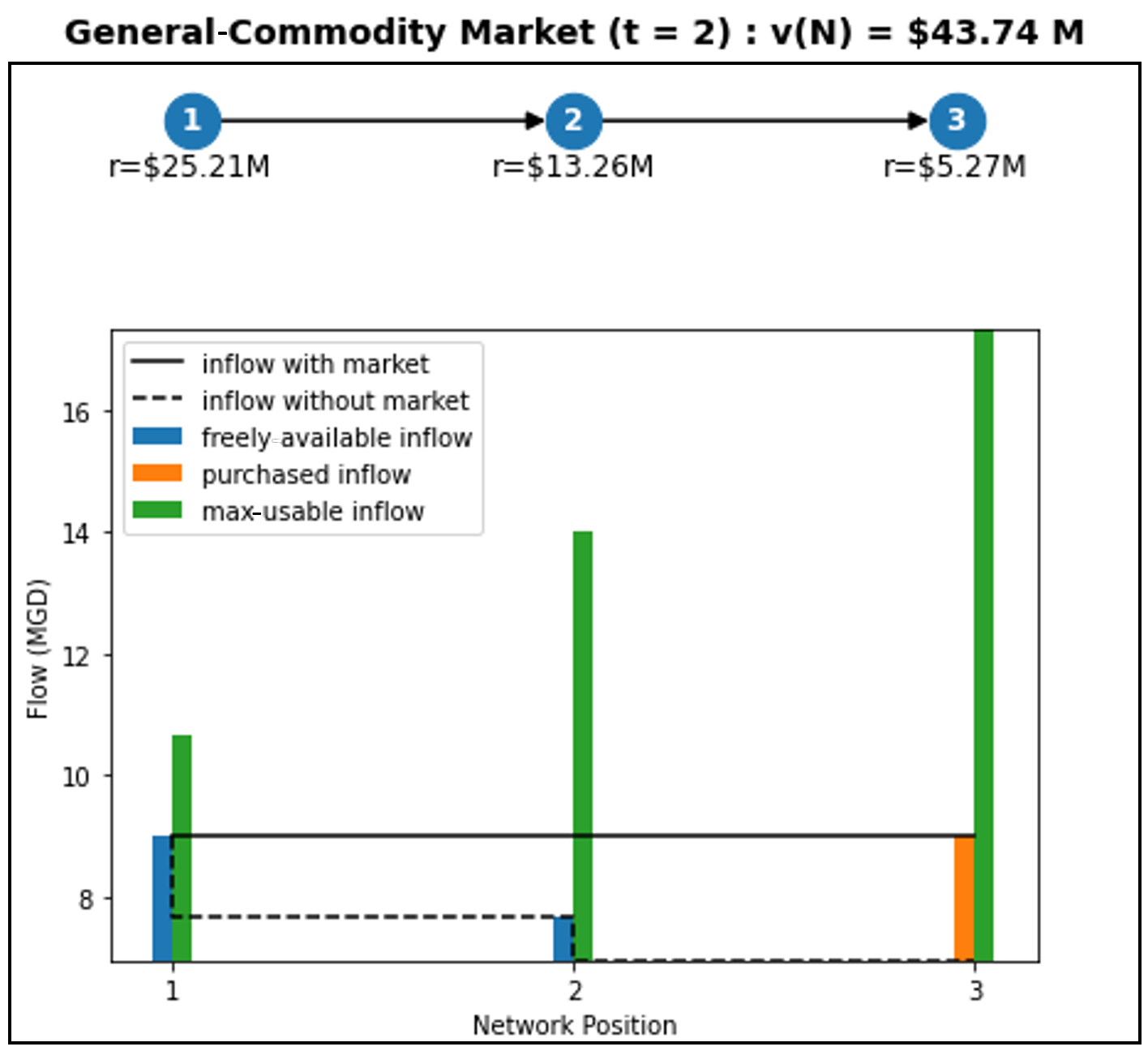}
        \caption{General commodity market with purchase protection.}
        \label{fig:S113_GCM}
    \end{subfigure}
    \begin{subfigure}[b]{0.4\paperwidth}
        \centering
        \includegraphics[width=\linewidth]{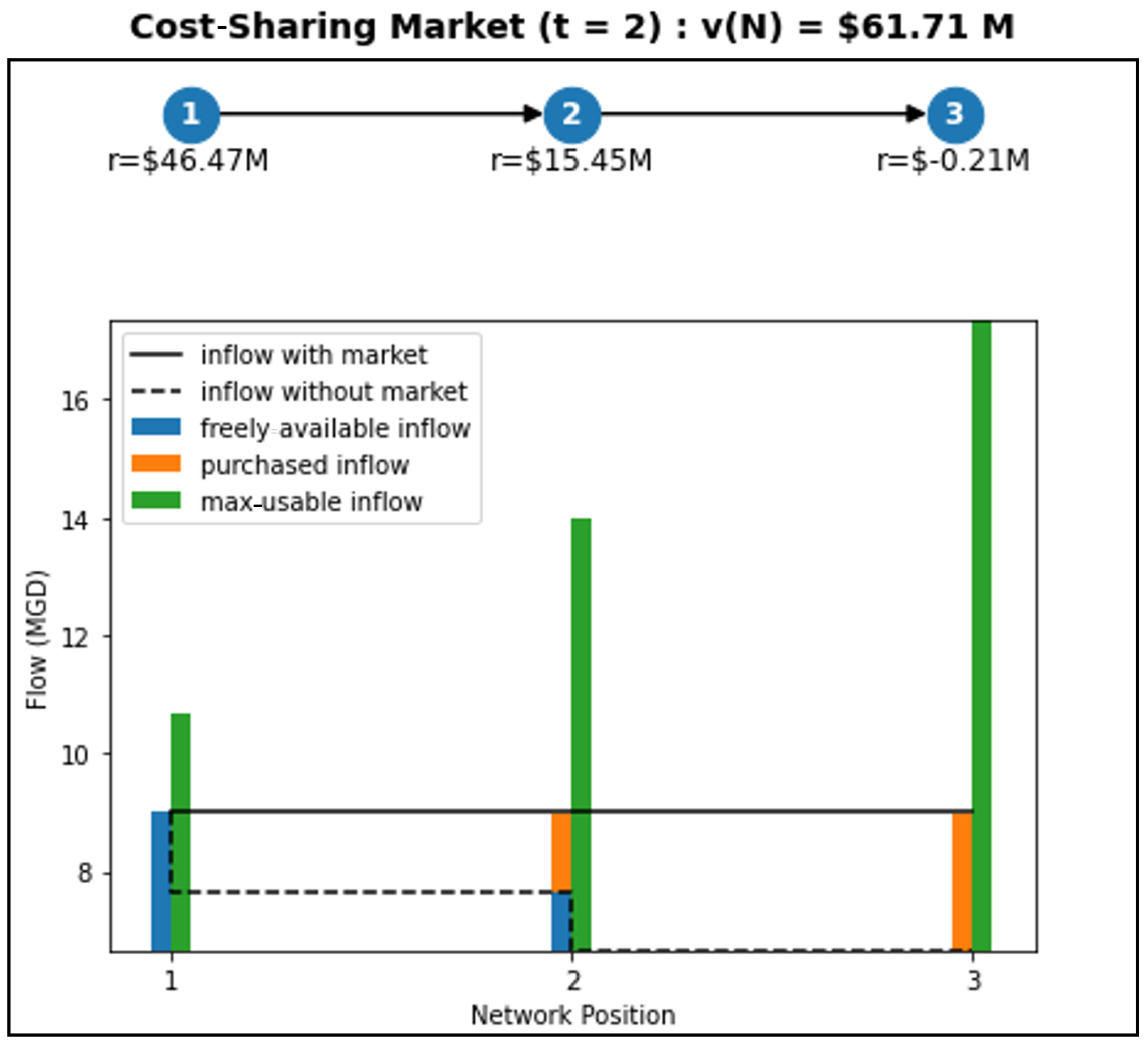}
        \caption{Cost-sharing market with intermediate positional advantage.}
        \label{fig:S113_CSM}
    \end{subfigure}
    \label{fig:S113main}
    \caption{Visualization of the Downstream Economic Growth Scenario. The cooperative game-theoretic rewards to each player, r, are shown in the plan view of the network. Water usage is shown in the profile view.}
\end{figure}

Figures \ref{fig:S47_GCM} and \ref{fig:S47_CSM} compare and contrast the market structures for the Large Prosumer Scenario, while Figures \ref{fig:S113_GCM} and \ref{fig:S113_CSM} do the same for the Downstream Economic Growth Scenario. All these figures depict the line graph of the river in plan view with the cooperative game theory metrics. Additionally, a profile view representing the flow levels in the river is depicted in millions of gallons per day (MGD). These plots have two key groups of elements warranting explanation.

The first group of elements includes the bars on the profile view. The stacked bar represents the actual inflow. It is subdivided into the freely-available inflow irrespective of the market structure plus additional inflow purchased from a market. The maximum usable inflow for each player is also plotted to gauge the level of scarcity of the actual inflow. In essence, it represents the inflow levels that the player would want to see in the river. 

The second group of elements includes the line plots on the profile view. They represent how consumptive losses and the subsequent reductions impact the actual flows on the river. Specifically, the inflow without market represents the minimum river levels at a player's node if no consumptive-loss reductions were made. By contrast, the inflow with market represents the actual levels in the river including consumptive-loss reductions. If these two line plots are equal, then no consumptive-loss reductions were made. For additional clarity, Table \ref{tab:det_an_def} mathematically defines these elements and other terms used in the detailed analysis.

\begin{table}[]
\centering
\begin{tabular}{|l|l|}
\hline
Term [Reference]                   & Mathematical Definition                                                                            \\ \hline
inflow with market (MGD) [D1]      & $n_i+O^{min}_{i-1,t}+\sum_{j=1}^{|I|}\sum_{c=1}^{|C|}\delta^{all}_{us_{j,i}}L^R_{i,c,t}+W^S_{j,t}$ \\ \hline
inflow without market [D2]  & $n_i+O^{min}_{i-1,t}$                                                                              \\ \hline
freely available inflow (MGD) [D3] & $Q_{i,t}-W^P_{i,t}+r^{fc}_{i,t}$                                                                   \\ \hline
max-usable inflow (MGD) [D4]     & $demand_{i,t}+r^{fc}_{i,t}$                                                                        \\ \hline
resource utilization (\%) [D5]  & $1 - 
\min\left\{{\frac{D1 - (Q_{i,t}+r^{fc}_{i,t})}{D1},\frac{D4 - (Q_{i,t}+r^{fc}_{i,t})}{D4}}\right\}$                       \\ \hline
\end{tabular}
\caption{Mathematical definitions for terms used in the detailed analysis.}
\label{tab:det_an_def}
\end{table}

In the Large Prosumer Scenario, the CSM generates higher social welfare than the GCM because the former has higher resource utilization than the latter. In both market structures, Player 2 purchases inflow up to the amount made available through the markets. However, as shown in Figure \ref{fig:S47_GCM}, Player 3's actual inflow in the GCM is less than the available inflow with the market. This difference between available and actual inflow used to meet demand represents resource under-utilization because all the available inflow with the market is within Player 3’s maximum usable inflow. By contrast, Figure \ref{fig:S47_CSM} shows that both Player's 2 and 3 are able to purchase all the available inflow in the CSM.

The resource under-utilization in the GCM is a consequence of the structure of the market-clearing conditions. In  \ref{eqn:gcm_mc}, water-release purchases (i.e., $W^P_{j,i,t}$) are treated as bilateral agreements between a supplier i and a purchaser j. These agreements will increase the actual inflow in the river for all players downstream of Player i unless Player j's consumptive losses are 100\%. However, the GCM structure only permits Player j to withdraw this added inflow. By contrast, the CSM relaxes the bilateral purchasing assumption (i.e., $W^P_{i,t}$) in  \ref{eqn:MC} to allow multiple players to benefit from the water releases. 

In the Downstream Economic Growth Scenario, the GCM generates higher social welfare than the CSM because the bilateral water purchases reduce the asymmetry in the network. As shown in Figure \ref{fig:S113_GCM}, the GCM decreases the loss reductions that Player 3 needs to purchase (2.07 MGD instead of 2.33 MGD) to maximize social welfare. This occurs because the water Player 3 purchases from Player 1 is inaccessible to Player 2. In the CSM, no such restriction exists, so Player 2 uses this additional water as shown in Figure \ref{fig:S113_CSM}. In doing so, Player 2 incurs additional losses because it has a non-zero loss factor.

This scenario illustrates that the overall decrease in social welfare from intermediate losses can be significant because Player 2 values an incremental increase in consumption less than Player 3. Numerically, Player 3 has a higher $\gamma^{flow}$ value than Player 2 (\$20.55 M vs. \$15.95 M), which represents the marginal value of additional water use. As alluded to earlier in this section, this phenomenon in the CSM market structure gives a positional advantage to intermediate players. Relative to the GCM, Player 2 increases water withdrawals and subsequently increases Player 3's consumptive loss-reduction purchases.

The Downstream Economic Growth Scenario also reveals an important nuance with regards to social welfare. The CSM model technically has a higher characteristic function value, but it is not an imputation because the reward to Player 3 is negative. Thus, the basin would not likely agree unanimously to implement this market structure. Alternatively, the GCM model is an imputation. Thus, the GCM is a more reasonable alternative for improving social welfare over the no-market structure.

In both scenarios, the figures for time period 1 were omitted because no purchases of loss reductions are made. This occurs because the players making loss reductions realize a higher price by waiting for demand increases. This can be observed mathematically in the stationary KKT conditions for loss reductions (Equations \ref{eqn:gcm_stat_LR} and \ref{eqn:csm_stat_LR}). The tendency to withhold supply may offset some of the inherent advantages of market efficiencies.

Table \ref{tab:mult_price} illustrates the non-unique solutions in the GCM market. The term "V" represents the starting point used for all the variables in the model. Thus, different solution starting points generate more than one valid solutions that satisfy Theorem \ref{th:prices2}. An equity-enforcing set of constraints found in discretely constrained MCPs would be one way to distinguish between multiple solutions\cite{djeumou2019applications, gabriel2017solving}.

\begin{table}[]
\centering
\begin{tabular}{c|c|c}
Variable   / Parameter                             & V = 2 & V = 2000 \\ \hline
$\gamma^{loss}_{2,1,1}$                            & 0.71  & 1.34     \\
$\gamma^{loss}_{2,1,2}$                            & 6.84  & 6.52     \\
$c^{cu}_{2,1,1}$                                   & 1     & 1        \\
$\gamma^{loss}_{2,2,1}$                            & 0     & 0.63     \\
$\gamma^{loss}_{2,2,2}$                            & 3.55  & 3.24     \\
$c^{cu}_{2,2,1}$                                   & 5     & 5        \\
$\sum_{t=1}^2\gamma^{loss}_{2,1,t}+c^{cu}_{2,1,1}$ & 8.55  & 8.87     \\
$\sum_{t=1}^2\gamma^{loss}_{2,2,t}+c^{cu}_{2,2,1}$ & 8.55  & 8.87     \\
$\pi^{as}_{2,1}$                                   & 8.55  & 8.87     \\
$\gamma^{flow}_{3,1}$                              & 8.55  & 8.87     \\
$W^P_{3,2,1}$                                      & 0.67  & 0.68    
\end{tabular}
\caption{Multiple Prices Scenario in the GCM Market}
\label{tab:mult_price}
\end{table}

\subsubsection{Duck River Model}
Another model was also generated for the Duck River in Tennessee. The purpose of this model is to evaluate the market approaches with real-world data and to explore the impact of capital projects. In the three-node model, no capital projects were considered. In contrast, the Duck River model seeks to understand how the water release market impacts the optimal timing of water infrastructure investments with large fixed costs. This use case is relevant to river basins in general because water scarcity usually serves as a driver for significant supply expansion. For context, a similar real-options approach was considered for a single-optimization problem in Chapter 18 of \cite{daniell2015understanding}.

Stakeholders in the Duck River watershed in central Tennessee have been navigating water resource-related challenges in earnest since the extreme drought of 2007. Seven water utilities in the Duck River watershed serve water to approximately 250,000 people and industries that include car manufacturers, food processing plants, and other businesses. In addition to these uses, the river provides a wide range of other values including recreation, an excellent fishery, and some of the most biologically-rich freshwater habitat in North America \cite{obg2011supply}.

The drought of 2007 highlighted the issue that in extended dry weather conditions, the citizens of the Duck River region depend on the water stored in Normandy Reservoir to meet all designated uses, including drinking water, wastewater assimilation, recreation, and natural resource protection. The dramatic decrease in rainfall, combined with the multitude of uses of the reservoir and the river, caused record low water levels in Normandy Reservoir that resulted in temporary changes in dam operation to protect water uses. Weather patterns and growth projections, combined with the obligation to manage water resources responsibly for future generations, created the need for a comprehensive regional water supply plan for the Duck River Region.\cite{obg2011supply}

Development of the regional water supply plan \cite{obg2011supply} and a reservoir-river model of water budgets along the river highlighted the inequities in benefits between the upstream and downstream users in the basin \footnote{The OASIS software was used to build the reservoir-river model. More information can be found here: www.hazenandsawyer.com/publications/oasis-modeling-for-water-people/}. Higher than normal releases from the reservoir to the river during extreme dry events resulted in impacts to the reservoir users due to excessive draw down of the reservoir. In contrast, the flow constraint of 100 cubic feet per second (cfs) on just the most downstream user ignored consumptive water use by upstream water systems, golf courses, and other users.  

While the river-reservoir model provided insights into water withdrawals and discharges along the river (i.e., water balance), improved decision-making tools are needed to gain a better understanding of the following questions:
\begin{itemize}
    \item How does the flow of water in the Duck River translate into the flow of economic and environmental benefits for the individual stakeholders and the region?
    \item How can decision-makers overcome the strategic fragmentation that exists among stakeholders who individually may have an incomplete view of the “big-picture” problems for the region?
    \item How can water supply agreements and permits incorporate flexibility to overcome changing conditions?
\end{itemize}

\begin{figure}
\centering
\includegraphics[scale = 0.75]{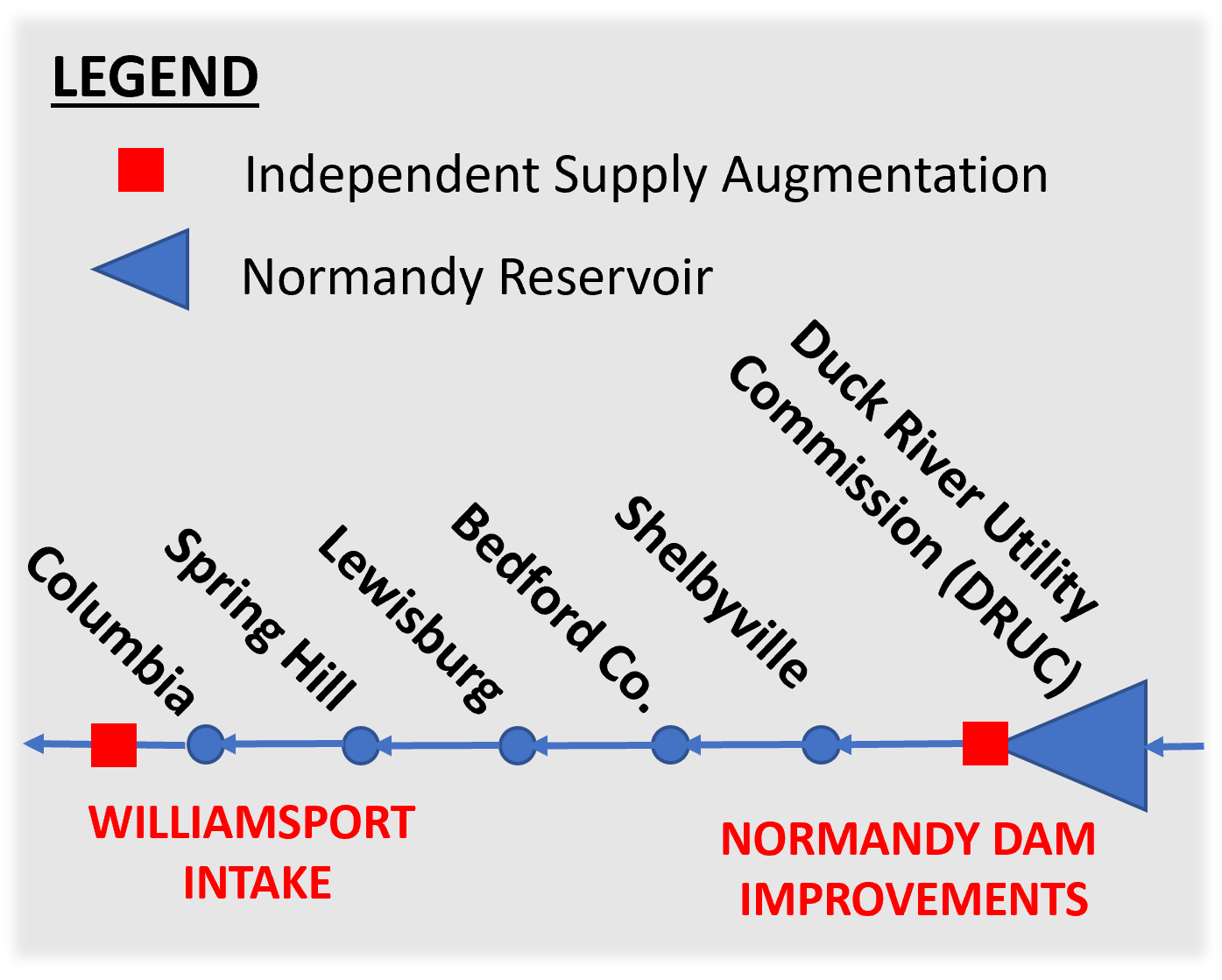}
\caption{Schematic of the Duck River Agency's municipal users and relative positions along the river. Normandy reservoir is at the upstream end of the river. The flow direction is shown opposite of convention to reflect the east-to-west flow of the river.}
\label{fig:DRA_schem}
\end{figure}

Figure \ref{fig:DRA_schem} depicts the line-graph network structure for the Duck River Basin. Contrasting with the previous example, there are six players considered instead of three. Each player is a municipal water provider for the named city or county with the exception of Duck River Utility Commission (DRUC). DRUC serves the cities of Manchester and Tullahoma using the water from Normandy reservoir at the upstream end of the basin. The Normandy dam separates DRUC from the rest of the downstream water providers.

Several data sources were consulted to estimate the model parameter values. The published water rates for each of the municipalities were used to estimate water operating costs. The water supply plan \cite{obg2011supply} and the drought management plan \cite{obg2013drought} were used to estimate water supply volumes, capital recovery costs, capital supply augmentation capacities, and regulatory flow constraints. Water demand projections and consumptive loss fractions were estimated from analysis associated with a demand projection study in the basin \cite{maddaus2016demand}. Industry-available data was used to estimate consumptive use costs and the inverse demand slope \cite{alcubilla2006derived} \cite{pickard2007reducing} \cite{Ramboll_WS2020}. 

The primary water supply source for most of the basin is water released from Normandy reservoir. Historically, these releases have been at least 77.6 MGD at a 97 percent reliability \cite{obg2013drought} \footnote{This is based on the criteria for Stage 3 drought conditions, which represents the drought threshold necessitating the reduction of Normandy Dam outflows to 77.6 MGD. This condition is expected to occur once every thirty years.}. Due to rapid growth in the region, water demands are expected to increase in the coming decades. Furthermore, consumptive use represents over a quarter of the water withdrawn from the basin in some cases \cite{maddaus2016demand}. Some users discharge wastewater outside the basin as well. 

This supply and demand profile of the basin creates the potential for inequities in economic growth. Columbia is expected to grow from 7.54 to 12.74 MGD between the years 2015 and 2050. Despite the growth potential, Columbia has the least favorable access to water. First, it is the only water system with a regulatory flow constraint in its water withdrawal permit. The permit states that Columbia must cease all water withdrawals if the flow in the river decreases to 64.6 MGD. Additionally, it is the farthest downstream player, so the consumptive losses of upstream water systems can exacerbate Columbia's inequity.

The water supply plan for the Duck River Agency \cite{obg2011supply} identified two projects as alternatives to address this and other water supply related inequities in the basin. Figure \ref{fig:DRA_schem} depicts the location of these proposed capital projects in the basin. The Normandy Dam improvements project would raise the Normandy dam to effectively increase the water stored in the Normandy Reservoir. Alternatively, the Williamsport project would create a new water intake at a less environmentally-sensitive location, which would allow for increased withdrawals from the river. 

The latter alternative is incorporated into the present analysis because it has emerged as the higher-priority project for Columbia and the rest of the basin's stakeholders. The project enables Columbia to withdraw water without its regulatory flow constraint, which therefore improves its access to water significantly. It also decreases its dependence on releases from the Normandy reservoir, which leaves more water available for upstream players. Thus, the scenarios considered for this model involve the installation year for the Williamsport Intake project. 

Assuming Columbia finances the project, the optimal timing for the investment is the key question underlying the scenario analysis in this model. Deferring investment decreases project costs. However, the decreased costs must be weighed against the benefits of the water supply increase. Furthermore, the role of the water release market must also be considered.

To answer this question, separate model runs for each market structure were performed for the installation years 2015,2020,..,2040,2045. It was assumed that Columbia bore the majority of the direct benefits and associated costs. Accordingly, the required augmentation parameter, $a^{req}_{columbia,t}$, and the capital cost parameter,$c^{cap}_{columbia,t}$, were manipulated from scenario to scenario. The former was set to the project's capacity, which is 64.6 MGD, for the installation year onward. The latter was set according to the simple formula: $c^{cap}_{columbia,t} = 5*[annual\_payment] / 64.6$. This converted the annual payment to cost per MGD per planning period. 

\begin{figure}
\centering
\includegraphics[scale = 0.7]{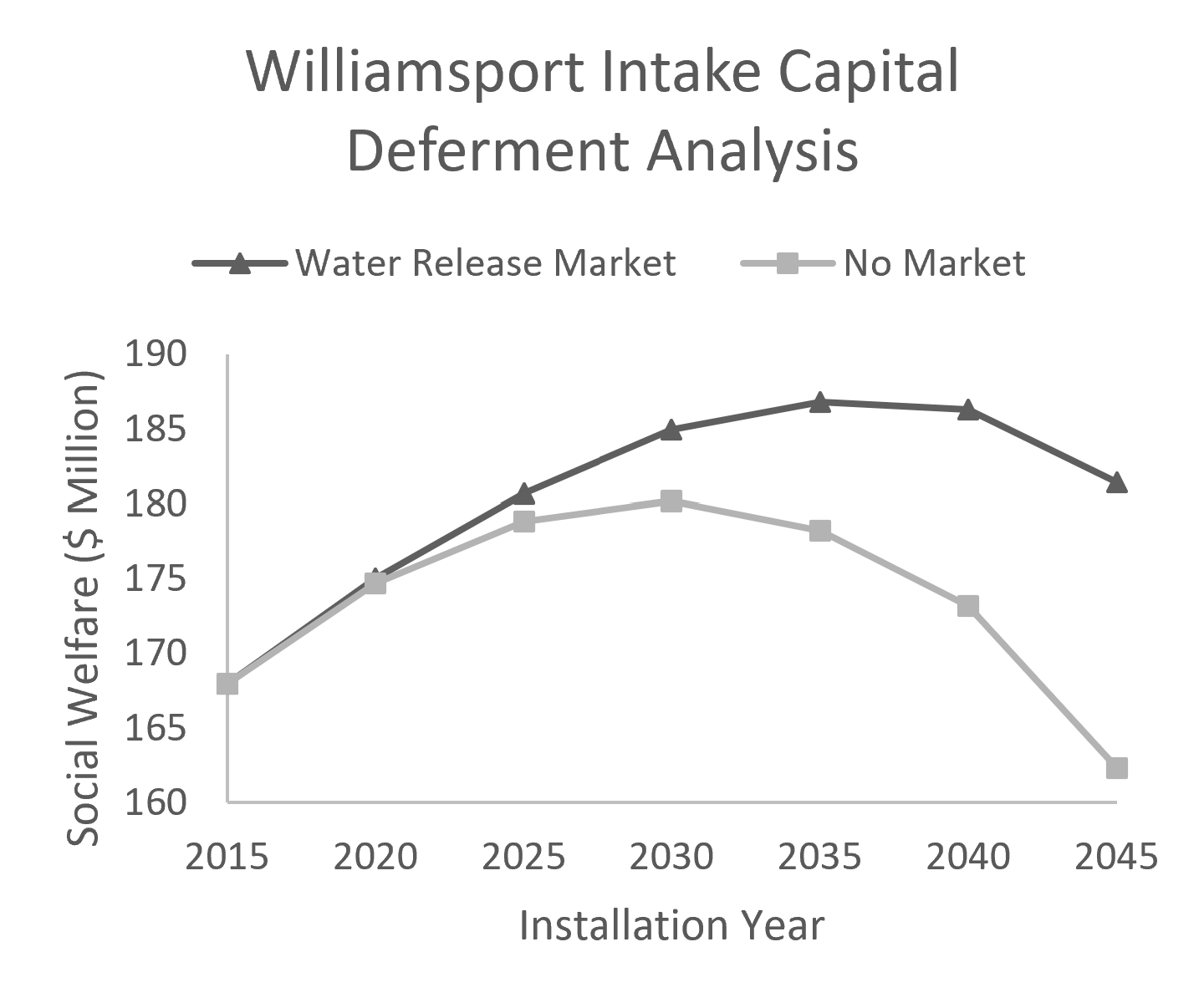}
\caption{Two plots of the social welfare for various installation years of the Williamsport Intake capital project for a water release market and no market.}
\label{fig:WI_Scenarios}
\end{figure}

Figure \ref{fig:WI_Scenarios} depicts the results of these scenarios. For each scenario, social welfare is plotted against the installation year for both the water release market and the no market models. The water release market is referred to generically because the GCM and CSM models produce identical results. This arises because there are no intermediate players purchasing water. Also, based on Theorem \ref{th:prices}, we see that the prices for all upstream nodes of Columbia active in the loss-reduction market must have similar nodal prices (in the GCM formulation).  Take for example the year 2025, in that case, Columbia purchases from DRUC, Shelbyville, and Spring Hill and the resulting, common market price for water is 0.874. The curves reveal competing trade-offs in the installation year for the project. Furthermore, these competing trade-offs occur at higher social welfare (i.e., benefits minus costs) for the water release market than the no market model. The water release market also enables the project to be deferred for longer. 
 As noted previously, the cost of the installation decreases the longer it is deferred. However, the cumulative opportunity costs, as measured through $\gamma^{flow}$, increase with deferment because less water is available to satisfy economic growth. These competing costs are jointly minimized between the years 2035 and 2040 for the water release market. 

\begin{figure}[htbp]
\centering
\includegraphics[scale = 1]{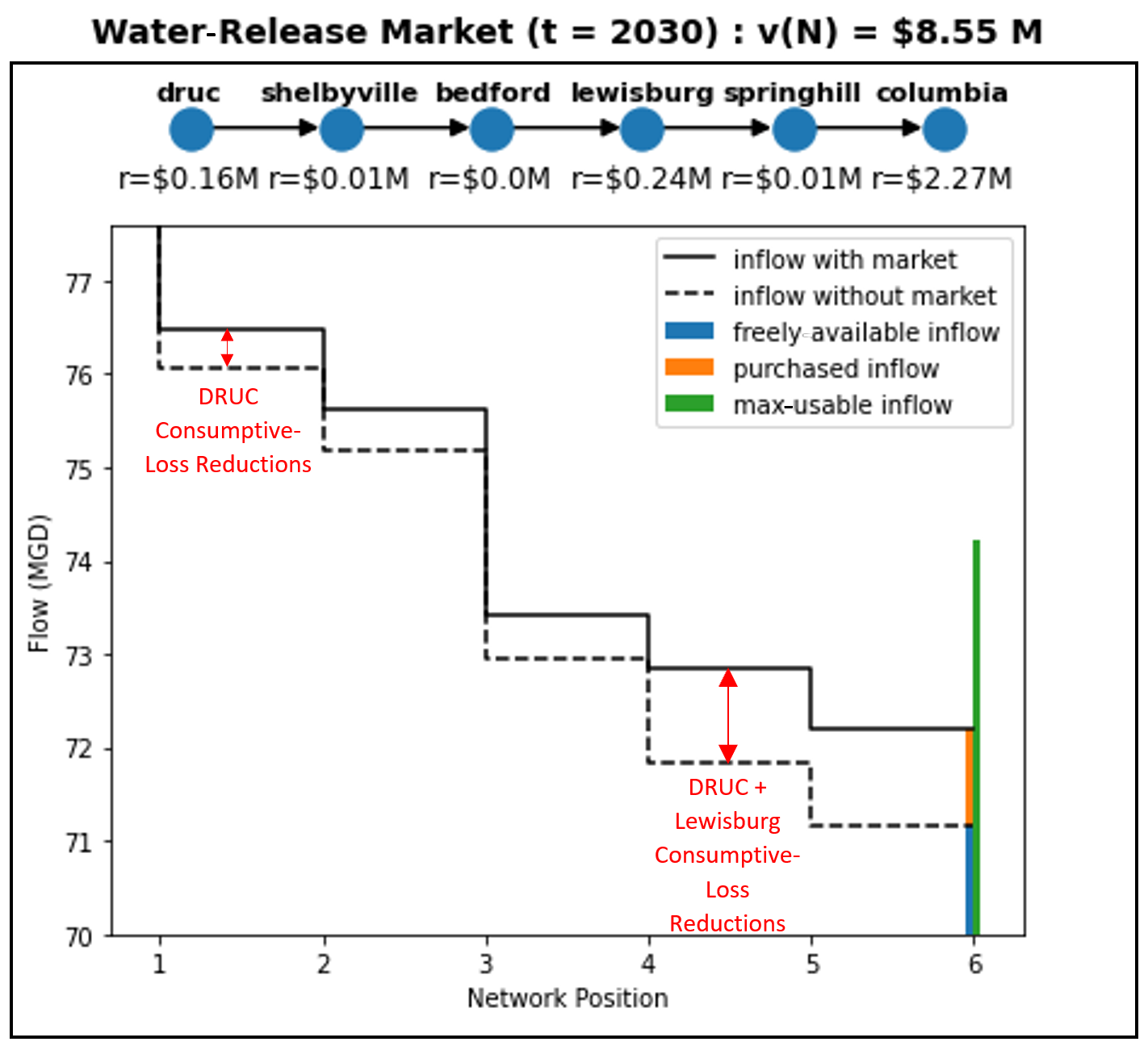}
\caption{This figure depicts the function of the water release market in 2030, which is five years before the optimal installation of the capital project.}
\label{fig:DRA_2030}
\end{figure}

The water release market decreases the water opportunity cost to make continued capital deferment viable, thereby serving as a temporary water supply solution. This can be visualized in Figure \ref{fig:DRA_2030}, which depicts the estimated flow conditions five years prior to the optimal time frame for the Williamsport Intake installation. The inflow needed to meet water demand is only a limiting factor (i.e., binding constraint) for Columbia because they are the only utility with a non-zero value for $r^{fc}_{i,t}$. To reduce the supply deficit, Columbia purchases a small amount of consumptive-loss reductions from each player to decrease the opportunity cost of deferring the Williamsport Intake project another five years. This can be observed in the differences between the "net inflow" and "min inflow" curves.

The Duck River case study reveals a potential unintended consequence of the water release market. Market profits (i.e.,$\pi^{as}_{i,t}-c^{cu}_{i,c,t}$) tend to increase water consumption beyond what it would be otherwise. As shown in Table \ref{tab:DRAWaterEfficiency}, each of the players upstream of Columbia occasionally use more water than their nominal demand during the Williamsport Intake deferment period (i.e., 2015-2030). Such excess water consumption only occurs when $\lambda^{sup}_{i,t}$ is positive, which can be interpreted as market profits.

Thus, increasing consumption allows the upstream users to experience more consumptive losses and subsequently derive more consumptive loss-reduction revenue from Columbia. All the upstream players that make loss reductions have a positive value for $\lambda^{sup}$ in 2030. This is when Columbia's water demand opportunity cost is the highest in the deferment period. Several players also have positive values for $\lambda^{sup}$ in 2015 because consumptive losses occur at lower rates for these players in later time periods. 

\begin{table}
\centering
\begin{tabular}{l|l|l|l|l}
Utility (i) & Time Period (t) & $Q_{i,t}$ (MGD) & Nominal Demand (MGD) & $\lambda^{sup}_{i,t}$ (\$ M) \\ \hline
DRUC        & 2030            & 7.5             & 7.27                 & 0.06                         \\
Shelbyville & 2015            & 3.98            & 3.85                 & 0.07                         \\
Shelbyville & 2030            & 5               & 4.89                 & 0.03                         \\
Lewisburg   & 2015            & 2.59            & 2.47                 & 0.06                         \\
Lewisburg   & 2030            & 3.27            & 3.17                 & 0.03                         \\
Spring Hill & 2015            & 2.79            & 2.66                 & 0.06                         \\
Spring Hill & 2030            & 3.95            & 3.84                 & 0.03                        
\end{tabular}
\caption{This table demonstrates how profit generated from the water release market (i.e., $\lambda^{sup}_{i,t}$) can increase water consumption. The nominal demand represents the projected water needs, and Q represents the total water withdrawn for consumption.}
\label{tab:DRAWaterEfficiency}
\end{table}

\section{Conclusions}
The river basin equilibrium formulations presented reveal a general framework for non-cooperative models on line graphs. In the three-node example, a market structure leads to non-negative rewards for all players. However, these models are needed to determine which market structure is necessary to achieve these rewards. In the Duck River example, the water-release market acts as a temporary solution to help optimally defer capital investments. In all cases, it was assumed that consumptive-loss reductions are practical and relatively cheap to implement. Water releases from storage become much more important when this assumption fails.

Future research will build on the water-release market approach to make it more versatile for more types of river basins. For example, the models in this paper assume that upstream players cannot completely control the water resource. This assumption could fail in situations with a few number of players or restricted water access. In such a case, treating the river basin as a Stackelberg game is perhaps more appropriate. An alternate bi-level formulation could consider the top level player as a government entity that imposes regulatory flow constraints on the lower level players. Lastly, considering the role of groundwater could be important in river basins with high agricultural usage.

\section*{Acknowledgements}
This research was funded by the National Science Foundation's Civil Infrastructure Systems Program (NSF Award \# 2113891). Aside from funding, the agency did not play any other role in the study design, data collection, analysis, interpretation, writing of the report, or publication decisions.

Thank you to Doug Murphy, the executive director of the Duck River Agency, for providing technical assistance, data, and advice featured in this paper.
\section*{Declaration of Interest}
There are no interests to declare.

\printbibliography

\section{Appendix}

\subsection*{Proof of Theorem \ref{th:imputation}}
\begin{proof} 
In this game, $r_i$ equals the difference between the optimal objective functions of the Generalized Nash reformulation and the original line-graph game:
\begin{equation}
    r_i = f^{GN}_i(x^*_i,y^*_i,z^*_i,\pi) - f^{LG}_i(x^*_i)
    \label{eqn:rewards}
\end{equation}
Assume $r_i \ge 0 \quad \forall i \in I$. In this case, $f^{GN}_i(x^*_i,y^*_i,z^*_i,\pi) = f^{LG}_i(x^*_i) \quad \forall i \notin I^C$ using an optimality argument. Otherwise, player i would chose to be in the set $I^C$. Therefore, $v(I)=v(I^C)=\sum_{i \in I^C}r_i$ follows from substituting Equation \ref{eqn:rewards} into Equation \ref{eqn:line_graph_cf} given that $r_i=0$ for all non-participating players. This satisfies Equation \ref{eqn:imput_cond_1}. The premise satisfies Equation \ref{eqn:imput_cond_2}.
\end{proof}

\subsection*{Proof of Theorem \ref{eq:existence}}

\begin{proof} \label{xxx}
First note that the KKT conditions to (\ref{eq:short}) are necessary due to the linearity of the constraint functions and sufficient due to the concavity of the objective function since $\beta_i>0$.  These conditions are the following.
\begin{subequations} \label{eq:short_all}
\begin{equation} \label{eq:short1}
    0 \leq c^{ops}_{i}-\alpha_i+\beta_iQ_i-lf_i\gamma^{loss}_i+\gamma^{flow}_i \perp Q_i \ge 0
\end{equation}
\begin{equation}\label{eq:short2}
0 \leq c^{cu}_i -\pi^{as}_i + \gamma^{loss}_i \perp L^R_i \geq 0   
\end{equation}
\begin{equation} \label{eq:short3}
0 \leq \pi^{as}_j -\gamma^{flow}_{i} \perp W^P_{ij} \geq 0, \forall j \in U_i
\end{equation}
\begin{equation} \label{eq:short4}
0 \leq lf_i Q_i-L^R_i \perp \gamma^{loss}_i \geq 0  
\end{equation}
\begin{equation}\label{eq:short5}
0 \leq n_i + \sum_{j \in U_i} W^P_{ij} -r^{fc}_i -Q_i + O^{min}_{i-1} \perp \gamma^{flow}_i \geq 0   
\end{equation}
\end{subequations}
Consider a node $e$ that is not active in the loss-reduction market.  Then, the following are feasible values for the variables:
\begin{equation*}
Q_e=0, L^R_e=0, W^P_{ej}=0, \forall j \in U_e, \gamma^{loss}_e=0, \gamma^{flow}_e=0, \pi^{as}_e=0. 
\end{equation*}
To see why, take each part of (\ref{eq:short_all}) separately. (\ref{eq:short1}) is feasible as long as $c^{ops}_e -\alpha_e \geq 0$, which is condition (ii) in the premise.  (\ref{eq:short2}) holds as long as $c^{cu}_e \geq 0$ which is guaranteed since all cost coefficients are positive. Conditions (\ref{eq:short3}),  (\ref{eq:short4}) and the market-clearing condition (\ref{eq:MCC_short}) are automatically satisfied for the given values.  Note that for $j\in U_e$,  (\ref{eq:short3}) holds if $\pi^{as}_j >0$ or if $\pi^{as}_j=0$. Under (iii), and considering the definition of $O^{min}_e$, we see that from  (\ref{eqn:omin_short}) $O^{min}_e=\sum_{r=1}^{e} n_r$.  Thus, in (\ref{eq:short5}), we need $n_e +O^{min}_{e-1}=\sum_{r=1}^{e} n_r \geq r^{fc}_e$ which is (iv).

Now consider node $u$ which alone supplies loss reduction. As $L^R_u >0$ this means by (\ref{eq:short4}) that $Q_u>0$. $L^R_u>0$ in (\ref{eq:short2}) means that $\pi^{as}_u=c^{cu}_u+\gamma^{loss}_u>0$. This in turn implies in (\ref{eq:MCC_short}) and (iii) that $W^P_{du}=L^R_u=lf_u Q_u>0$ is feasible.  Since $Q_u>0$, choose  $\gamma^{loss}_u=\gamma^{flow}_u=0$ so that in  (\ref{eq:short1}) $Q_u=\frac{\alpha_u-c^{ops}_u}{\beta_u}$ which is positive by the left-most inequality in (v).  (\ref{eq:short3}) is true since $\forall j \in U_u$, $W^P_{uj}=0$, $0=\gamma^{flow}_u \leq \pi^{as}_j=0$. Note that since $lf_uQ_u=L^R_u$ and $\gamma^{loss}_u \geq 0$, (\ref{eq:short4}) is feasible.  (\ref{eq:short5}),  reduces to
\begin{equation*}
0<Q_u\leq \sum_{r=1}^u n_r -r^{fc}_u \perp \gamma^{flow}_u \geq 0
\end{equation*}
which is true as long as $Q_u$ satisfies
\begin{equation*}
0<Q_u=\frac{\alpha_u-c^{ops}_u}{\beta_u} \leq \sum_{r=1}^u n_r -r^{fc}_u
\end{equation*}
which is valid by the right-most inequality in (v).  

Lastly, consider node $d$, a sole downstream node in $D_u$, active in the loss-reduction market and receiving water purchases from node $u$.  By construction, from (\ref{eq:short3}), $W^P_{du}>0 \Rightarrow \gamma^{flow}_d=\pi^{as}_u>0$, the right-most  part coming from the analysis above.  Now let the rest of the variables have the following values: $\gamma^{loss}_d =0, Q_d=0, L^R_d=0, \pi^{as}_d=0, \gamma^{flow}_d=\pi^{as}_u >0, W^P_{dj}=0, j \neq u, W^P_{dj}=L^R_u>0, j=u$. 
With those values, note that (\ref{eq:short1}) is satisfied as long as $\pi^{as}_u\geq \alpha_d-c^{ops}_d$ which is guaranteed by condition (vi) and taking into account the analysis above that gave $\pi^{as}_u=c^{cu}_u+\gamma^{loss}_u=c^{cu}_u$.    (\ref{eq:short2}) is automatically satisfied with those values since $c^{cu}_d \geq 0$. (\ref{eq:short3}) is satisfied since $W^P_{du}>0 \Rightarrow \gamma^{flow}_{d}=\pi^{as}_u$. (\ref{eq:short4}) is automatically satisfied for the given values as is (\ref{eq:MCC_short}), the latter since $W^p_{du}=L^R_u$ and $\pi^{as}_u>0$.  Since $\gamma^{flow}_{d}>0$, and given that $W^p_{du}=L^R_u=Q_u lf_u$ the last condition, (\ref{eq:short5}) is satisfied as long as:
\begin{equation*}
Q_d=\sum_{r=1}^d n_r-r^{fc}_d+\left(\frac{\alpha_u-c^{ops}_u}{\beta_u} \right) lf_u\\
\end{equation*}

which is guaranteed by (vii).

\end{proof}

\subsection*{Proof of Theorem \ref{th:prices}}
\begin{proof}
 Consider a node $j \in U_{it}^+$.  Since $W^P_{ijt}>0$ it follows by complementarity that $\pi^{as}_{j,t}=\frac{\gamma^{flow}_{i,t}}{d_t}$.  As this is true $\forall j  \in U_{it}^+$ we see that the corresponding prices $\pi^{as}_{j,t}$ must be equal with $\pi_{it}=\frac{\gamma^{flow}_{i,t}}{d_t}$.
\end{proof}

\subsection*{Proof of Theorem \ref{th:prices2}}
\begin{proof}
Consider a node $i$ for which $L^R_{cit}>0$, for some $c \in C, t \in T$.  Then, by (\ref{eqn:gcm_stat_LR}), 
\begin{equation*} 
 \pi^{as}_{it} =\frac{\sum_{t'=t} ^{|T|} \gamma^{loss}_{i,c,t}}{d_t}+c^{cu}_{cit}   
\end{equation*}
Since $k \in D_i^+$, then by (\ref{eq:Wp}), $W^P_{k,i,t} >0$, we see that 
\begin{equation*} \label{eqn:part b}
 \pi^{as}_{it}=\frac{\gamma^{flow}_{k,t}}{d_t}, \forall k \in D_i^+    
\end{equation*}
\end{proof}

\subsection*{LCP for the GCM Formulation} 
   \begin{subequations}\label{eq:KKT_GCM}
        \begin{equation}
            0 \le \sum^{|T|}_{t'=t}\lambda^{sup}_{i,t'} - \sum^{|T|}_{t'=t}\sum^{|C|}_{c=1}lf_{c,i,t}\gamma^{loss}_{c,i,t'} \perp W^{D}_{i,t} \ge 0 \quad \forall t
            \label{eqn:gcm_stat_WD}
        \end{equation}
        \begin{equation}
            0 \le d_t(c^{sr}_{i,t}-\pi^{as}_{i,t})-\gamma^{flow}_{i,t}+\gamma^{cap}_{i,t} \perp W^{S}_{i,t} \ge 0 \quad \forall t \label{eqn:b}
        \end{equation}
        \begin{equation}
            0 \le d_t(c^{ops}_{i,t} - \theta_{i,t}(Q_{i,t})) - \lambda^{sup}_{i,t} + \gamma^{flow}_{i,t} \perp Q_{i,t} \ge 0 \quad \forall t
            \label{eqn:gcm_stat_Q}
        \end{equation}
        \begin{equation}
            0 \le d_tc^{cap}_{i,t} - \gamma^{cap}_{i,t} + \lambda^{aug}_{i,t}\perp K_{i,t} \ge 0 \quad \forall t \label{eqn:d}
        \end{equation}
        \begin{equation}
            0 \le d_t(c^{cu}_{i,c,t} - \pi^{as}_{i,t})+\sum^{|T|}_{t'=t}\gamma^{loss}_{c,i,t'} \perp L^{R}_{i,c,t}  \ge 0 \quad \forall c,t 
            \label{eqn:gcm_stat_LR}
        \end{equation}
        \begin{equation} \label{eq:Wp}
            0 \le \delta^{all}_{us_{j,i}}(d_t\pi^{as}_{j,t}-\gamma^{flow}_{i,t}) \perp W^{P}_{ij,t} \ge 0 \quad \forall j,t \quad , \quad j \neq i 
        \end{equation}
        \begin{equation}
            0 \le \sum^t_{t'=1}lf_{c,i,t'}W^D_{i,t'} - L^{R}_{c,i,t'} \perp \gamma^{loss}_{i,c,t} \ge 0 \quad \forall c,t \label{eqn:g}
        \end{equation}
        \begin{equation}
            0 \le n_{i} + W^S_{i,t} + \sum^{|I|}_{j=1}\delta^{all}_{us_{j,i}}W_{ij,t}^{P} - r^{fc}_{i,t} +
            O^{min}_{i-1,t}-Q_{i,t} \perp \gamma^{flow}_{i,t} \ge 0 \quad \forall t \label{eqn:h}
        \end{equation}
        \begin{equation}
            0 \le K_{i,t}-W^{S}_{i,t} \perp \gamma^{cap}_{i,t} \ge 0
            \quad \forall t \label{eqn:i}
        \end{equation}
        \begin{equation}
             \sum^t_{t'=1}W^D_{i,t'} - Q_{i,t} = 0 \quad , \quad \lambda^{sup}_{i,t} \quad free \quad \forall t  \label{eqn:i}
        \end{equation}
        \begin{equation}
            K_{i,t} - a^{req}_{i,t} = 0 \quad , \quad \lambda^{aug}_{i,t} \quad free
            \quad \forall t \label{eqn:k}
        \end{equation}
        \begin{equation}
            O^{min}_{i,t} = n_{i}-\sum^{|C|}_{c=1}\sum^t_{t'=1}lf_{c,i,t'}W^D_{i,t'}+\sum^{|C|}_{c=1}\sum^{t-1}_{t'=1}L^{R}_{c,i,t'}+O^{min}_{i-1,t} \quad \forall i,t \label{eqn:omin}
        \end{equation}
        \begin{equation}
                \sum^{|I|}_{k=1}\delta^{all}_{ds_{k,i}} W_{ki,t}^{P} \leq  \sum_{c \in C} L^{R}_{i,c,t}+W^{S}_{i,t} \perp \pi^{as}_{i,t} \geq 0 , \forall t \label{eqn:MC1}
        \end{equation}
        \begin{equation}
            \theta_{i,t}(Q_{i,t}) = \alpha_{i,t} - \beta_{i,t}Q_{i,t} \quad \forall t \label{eqn:theta}
        \end{equation}
    \end{subequations}
    The endogenous functions were substituted into Equation (\ref{eqn:objfn}) prior to solving for the KKT conditions.

\subsection*{LCP for the CSM Formulation}
  \begin{subequations}
        \begin{equation}
            0 \le \sum^{|T|}_{t'=t}\lambda^{sup}_{i,t'} - \sum^{|T|}_{t'=t}\sum^{|C|}_{c=1}lf_{c,i,t}\gamma^{loss}_{c,i,t'}\perp W^{D}_{i,t} \ge 0 \quad \forall t
        \end{equation}
        \begin{equation}
            0 \le d_t(c^{sr}_{i,t}-\sum^{|I|}_{k=1}\pi^{as}_{k,t}\delta^{all}_{ds_{k,i}})-\gamma^{flow}_{i,t}+\gamma^{cap}_{i,t} \perp W^{S}_{i,t} \ge 0 \quad \forall t
        \end{equation}
        \begin{equation}
            0 \le d_t(c^{ops}_{i,t} - \theta_{i,t}(Q_{i,t})) - \lambda^{sup}_{i,t} + \gamma^{flow}_{i,t} \perp Q_{i,t} \ge 0 \quad \forall t
        \end{equation}
        \begin{equation}
            0 \le d_tc^{cap}_{i,t} - \gamma^{cap}_{i,t} + \lambda^{aug}_{i,t}\perp K_{i,t} \ge 0 \quad \forall t
        \end{equation}
        \begin{equation}
            0 \le d_t(c^{cu}_{i,c,t} - \sum^{|I|}_{k=1}\pi^{as}_{k,t}\delta^{all}_{ds_{k,i}})+\sum^{|T|}_{t'=t}\gamma^{loss}_{c,i,t'}\perp L^{R}_{i,c,t}  \ge 0 \quad \forall c,t
            \label{eqn:csm_stat_LR}
        \end{equation}
        \begin{equation}
            0 \le d_t\pi^{as}_{i,t}-\gamma^{flow}_{i,t}\perp W^{P}_{i,t} \ge 0 \quad \forall t
        \end{equation}
        \begin{equation}
            0 \le \sum^t_{t'=1}lf_{c,i,t'}W^D_{i,t'} - L^{R}_{c,i,t'} \perp \gamma^{loss}_{i,c,t} \ge 0 \quad \forall c,t
        \end{equation}
        \begin{equation}
            0 \le n_{i} + W^S_{i,t} + W^{P}_{i,t} - r^{fc}_{i,t} +
            O^{min}_{i-1,t}-Q_{i,t} \perp \gamma^{flow}_{i,t} \ge 0 \quad \forall t 
        \end{equation}
        \begin{equation}
            0 \le K_{i,t}-W^{S}_{i,t} \perp \gamma^{cap}_{i,t} \ge 0
            \quad \forall t
        \end{equation}
        \begin{equation}
             \sum^t_{t'=1}W^D_{i,t'} - Q_{i,t} = 0 \quad , \quad \lambda^{sup}_{i,t} \quad free \quad \forall t 
        \end{equation}
        \begin{equation}
            K_{i,t} - a^{req}_{i,t} = 0 \quad , \quad \lambda^{aug}_{i,t} \quad free
            \quad \forall t
        \end{equation}
        \begin{equation}
            O^{min}_{i,t} = n_{i}-\sum^{|C|}_{c=1}\sum^t_{t'=1}lf_{c,i,t'}W^D_{i,t'}+\sum^{|C|}_{c=1}\sum^{t-1}_{t'=1}L^{R}_{c,i,t'}+O^{min}_{i-1,t} \quad \forall t 
        \end{equation}
        \begin{equation} \label{eq:pi}
            \sum^{|I|}_{j=1}\delta^{all}_{us_{j,i}}(\sum^{|C|}_{c=1}L^{R}_{j,c,t}+W^{S}_{j,t})-W^{P}_{i,t} \ge 0 \perp \pi^{as}_{i,t} \ge 0 \quad \forall t
        \end{equation}
        \begin{equation}
            \theta_{i,t}(Q_{i,t}) = \alpha_{i,t} - \beta_{i,t}Q_{i,t}  \quad \forall t
        \end{equation}
    \end{subequations}
The endogenous functions were substituted into Equation (\ref{eqn:objfn}) prior to solving for the KKT conditions.

\subsection*{LCP for the No-Market Formulation}
\begin{subequations}
        \begin{equation}
            0 \le \sum^{|T|}_{t'=t}\lambda^{sup}_{i,t'} \perp W^{D}_{i,t} \ge 0 \quad \forall t
        \end{equation}
        \begin{equation}
            0 \le d_tc^{sr}_{i,t}-\gamma^{flow}_{i,t}+\gamma^{cap}_{i,t} \perp W^{S}_{i,t} \ge 0 \quad \forall t
        \end{equation}
        \begin{equation}
            0 \le d_t(c^{ops}_{i,t} - \theta_{i,t}(Q_{i,t})) - \lambda^{sup}_{i,t} + \gamma^{flow}_{i,t} \perp Q_{i,t} \ge 0 \quad \forall t
        \end{equation}
        \begin{equation}
            0 \le d_tc^{cap}_{i,t} - \gamma^{cap}_{i,t} + \lambda^{aug}_{i,t}\perp K_{i,t} \ge 0 \quad \forall t
        \end{equation}
        \begin{equation}
            0 \le n_{i} + W^S_{i,t}- r^{fc}_{i,t} +
            O^{min}_{i-1,t}-Q_{i,t} \perp \gamma^{flow}_{i,t} \ge 0 \quad \forall t 
        \end{equation}
        \begin{equation}
            0 \le K_{i,t}-W^{S}_{i,t} \perp \gamma^{cap}_{i,t} \ge 0
            \quad \forall t
        \end{equation}
        \begin{equation}
             \sum^t_{t'=1}W^D_{i,t'} - Q_{i,t} = 0 \quad , \quad \lambda^{sup}_{i,t} \quad free \quad \forall t 
        \end{equation}
        \begin{equation}
            K_{i,t} - a^{req}_{i,t} = 0 \quad , \quad \lambda^{aug}_{i,t} \quad free
            \quad \forall t
        \end{equation}
        \begin{equation}
            O^{min}_{i,t} = n_{i}-\sum^{|C|}_{c=1}\sum^t_{t'=1}lf_{c,i,t'}W^D_{i,t'}+O^{min}_{i-1,t} \quad \forall t
        \end{equation}
        \begin{equation}
            \theta_{i,t}(Q_{i,t}) = \alpha_{i,t} - \beta_{i,t}Q_{i,t}  \quad \forall t
        \end{equation}
    \end{subequations}

\end{document}